\pdfoutput=1
\documentclass[iop, apj, tighten]{emulateapj}
\usepackage{graphicx}
\usepackage[space]{grffile}
\usepackage{latexsym}
\usepackage{textcomp}
\usepackage{longtable}
\usepackage{multirow, booktabs}
\usepackage{amsfonts, amsmath, amssymb}
\usepackage{url}
\usepackage[utf8]{inputenc}
\usepackage{hyperref}
\hypersetup{colorlinks=false,pdfborder={0 0 0}}
\usepackage[english]{babel}
\usepackage{natbib}
\usepackage{enumitem}
\usepackage{color}

\newcommand{\msun}{\,\mbox{M}_\odot}

\newcommand{\kpc}{\,\mbox{kpc}}

\newcommand{\pc}{\,\mbox{pc}}

\newcommand{\gyr}{\,\mbox{Gyr}}
\newcommand{\myr}{\,\mbox{Myr}}

\newcommand{\yrinv}{\,\mbox{yr}^{-1}}
\newcommand{\kms}{\,\mbox{km}\,\mbox{s}^{-1}}

\newcommand{\mhalo}{M_{\rm halo}}

\newcommand{\mthm}{M_{200{\rm m}}}
\newcommand{\rthm}{R_{200{\rm m}}}

\newcommand{\mstar}{M_{\rm star}}

\newcommand{\sfr}{\mbox{SFR}}

\newcommand{\halpha}{\mbox{H}\alpha}

\newcommand\rmxaa{Rev. Mex. Astron. Astrofis.} % Revista Mexicana de Astronomia y Astrofisica
\newcommand\nar{New~Astron.~Rev.}    % New Astronomy Review

\shorttitle{Radial Migration in Low-Mass Galaxies}
\shortauthors{El-Badry et al.}

\begin{document}

%% LaTeX will automatically break titles if they run longer than
%% one line. However, you may use \\ to force a line break if
%% you desire.
\title{Breathing FIRE: How Stellar Feedback Drives Radial Migration, Rapid Size Fluctuations, and Population Gradients in Low-Mass Galaxies}

%% Use \author, \affil, and the \and command to format
%% author and affiliation information.
%% Note that \email has replaced the old \authoremail command
%% from AASTeX v4.0. You can use \email to mark an email address
%% anywhere in the paper, not just in the front matter.
%% As in the title, use \\ to force line breaks.

%\author{Kareem El-Badry}
%\affil{Yale University}
 
%\author{Andrew Wetzel}
%\affil{California Institute of Technology}
 
%\author{Marla Geha}
%\affil{Yale University}

\author{Kareem El-Badry\altaffilmark{1,2}, Andrew Wetzel\altaffilmark{2,3,6,7}, Marla Geha\altaffilmark{1}, Philip F. Hopkins\altaffilmark{2}, Dusan Kere{\v{s}}\altaffilmark{4}, \\ T. K. Chan\altaffilmark{4}, Claude-Andr{\'{e}} Faucher-Gigu{\`{e}}re\altaffilmark{5}}

\altaffiltext{1} {Department of Astronomy, Yale University, New Haven, CT, USA. kareem.el-badry@yale.edu}
\altaffiltext{2} {TAPIR, California Institute of Technology, Pasadena, CA USA}
\altaffiltext{3} {Carnegie Observatories, Pasadena, CA, USA}
\altaffiltext{4} {Department of Physics, Center for Astrophysics and Space Sciences, University of California at San Diego, La Jolla, USA}
\altaffiltext{5} {Department of Physics and Astronomy and CIERA, Northwestern University, Evanston, IL, USA}
\altaffiltext{6} {Moore Prize Fellow}
\altaffiltext{7} {Carnegie Fellow in Theoretical Astrophysics}
%% Notice that each of these authors has alternate affiliations, which
%% are identified by the \altaffilmark after each name.  Specify alternate
%% affiliation information with \altaffiltext, with one command per each
%% affiliation.

%\altaffiltext{1}{Visiting Astronomer, Cerro Tololo Inter-American Observatory.
%CTIO is operated by AURA, Inc.\ under contract to the National Science
%Foundation.}
%\altaffiltext{2}{Society of Fellows, Harvard University.}
%\altaffiltext{3}{present address: Center for Astrophysics,
%    60 Garden Street, Cambridge, MA 02138}
%\altaffiltext{4}{Visiting Programmer, Space Telescope Science Institute}
%\altaffiltext{5}{Patron, Alonso's Bar and Grill}

%% Mark off your abstract in the ``abstract'' environment. In the manuscript
%% style, abstract will output a Received/Accepted line after the
%% title and affiliation information. No date will appear since the author
%% does not have this information. The dates will be filled in by the
%% editorial office after submission.

\begin{abstract}
We examine the effects of stellar feedback and bursty star formation on low-mass galaxies ($\mstar=2\times10^6-5\times10^{10}\msun$) using the FIRE (Feedback in Realistic Environments) simulations. While previous studies emphasized the impact of feedback on dark matter profiles, we investigate the impact on the stellar component: kinematics, radial migration, size evolution, and population gradients. Feedback-driven outflows/inflows drive significant radial stellar migration over both short and long timescales via two processes: (1) outflowing/infalling gas can remain star-forming, producing young stars that migrate $\sim1\kpc$ within their first $100 \myr$, and (2) gas outflows/inflows drive strong fluctuations in the global potential, transferring energy to all stars. These processes produce several dramatic effects. First, galaxies' effective radii can fluctuate by factors of $>2$ over $\sim200\myr$, and these rapid size fluctuations can account for much of the observed scatter in radius at fixed $\mstar$. Second, the cumulative effects of many outflow/infall episodes steadily heat stellar orbits, causing old stars to migrate outward most strongly. This age-dependent radial migration mixes---and even inverts---intrinsic age and metallicity gradients. Thus, the galactic-archaeology approach of calculating radial star-formation histories from stellar populations at $z=0$ can be severely biased. These effects are strongest at $\mstar\approx10^{7-9.6}\msun$, the same regime where feedback most efficiently cores galaxies. Thus, detailed measurements of stellar kinematics in low-mass galaxies can strongly constrain feedback models and test baryonic solutions to small-scale problems in $\Lambda$CDM.
\end{abstract}

\bibliographystyle{apj}
%% Keywords should appear after the \end{abstract} command. The uncommented
%% example has been keyed in ApJ style. See the instructions to authors
%% for the journal to which you are submitting your paper to determine
%% what keyword punctuation is appropriate.

%% Authors who wish to have the most important objects in their paper
%% linked in the electronic edition to a data center may do so in the
%% subject header.  Objects should be in the appropriate "individual"
%% headers (e.g. quasars: individual, stars: individual, etc.) with the
%% additional provision that the total number of headers, including each
%% individual object, not exceed six.  The \objectname{} macro, and its
%% alias \object{}, is used to mark each object.  The macro takes the object
%% name as its primary argument.  This name will appear in the paper
%% and serve as the link's anchor in the electronic edition if the name
%% is recognized by the data centers.  The macro also takes an optional
%% argument in parentheses in cases where the data center identification
%% differs from what is to be printed in the paper.


%% From the front matter, we move on to the body of the paper.
%% In the first two sections, notice the use of the natbib \citep
%% and \citet commands to identify citations.  The citations are
%% tied to the reference list via symbolic KEYs. The KEY corresponds
%% to the KEY in the \bibitem in the reference list below. We have
%% chosen the first three characters of the first author's name plus
%% the last two numeral of the year of publication as our KEY for
%% each reference.
\section{Introduction}

Low-mass ``dwarf'' ($\mstar \lesssim 10^{9} \msun$) galaxies provide probes of structure formation on the smallest cosmological scales and thus represent compelling laboratories for testing the $\Lambda$CDM (cold dark matter plus a cosmological constant) framework. In addition, because low-mass galaxies reside in low-mass halos with shallow gravitational potential wells and low escape velocities, they are highly sensitive to stellar feedback as compared to more massive galaxies.

During episodes of star formation, radiation pressure, photoionization and photoelectric heating, stellar winds, and supernovae inject both energy and momentum into the interstellar medium (ISM). Especially in low-mass galaxies, these processes can fuel powerful galactic winds that drive significant ISM mass into the halo \citep{1974MNRAS.169..229L, 1986ApJ...303...39D, 2007A&A...474...67V} and cause the star formation rate (SFR) to decline until gas cools and reaccretes back into the galaxy. As a result, low-mass galaxies are thought to have highly stochastic and bursty SFRs \citep[for example,][]{1980ApJ...242..517G,2007ApJ...667..170S,2010ApJ...724...49M,2011ApJ...739....5W,Gonz_lez_Samaniego_2014}, which in turn can have significant effects on these galaxies' dynamical and morphological evolution. For instance, bursty star formation may explain the lack of coherent, rotationally-supported disks in galaxies with $\mstar \lesssim 10^{9.5} \msun$ \citep{2007MNRAS.382.1187K,Muratov_2015,2015MNRAS.452..986S} and the fact that star-forming galaxies are more likely to be dispersion-supported at low mass than at high mass \citep{2010MNRAS.406L..65S, 2010MNRAS.404L..60R, Teyssier_2013, 2015arXiv151101095W}.

Many studies have shown that stellar feedback-driven gas outflows can displace a significant fraction of a galaxy's total gas mass \citep[for example,][]{Mathews_1971, Haehnelt_1995, Muratov_2015, 2015arXiv150800007C}, and several recent works have investigated how the resulting time-varying potential can transfer orbital energy to dark matter. Baryon-driven potential fluctuations have been invoked as a possible solution to several of the discrepancies between the predictions of the $\Lambda$CDM framework and observation, including the ``core-cusp'' problem \citep[for example,][]{1996MNRAS.283L..72N,Mashchenko_2008,Pontzen_2012,O_orbe_2015,Chan_2015,2015arXiv150703590T} and the related ``too-big-to-fail'' problem \citep{Boylan_Kolchin_2011, Zolotov_2012, Garrison_Kimmel_2013, Chan_2015, 2015arXiv151200453D}, demonstrating that baryonic physics can have significant and lasting effects on the distribution of (collisionless) dark matter.

Stars are also (effectively) collisionless and thus feel the same fluctuations in the gravitational potential that transfer energy to dark matter. Therefore, one might expect the kinematics of stars to respond to feedback-driven outflows in much the same way as dark matter. Some theoretical works have shown that gas outflows/inflows can drive kinematic fluctuations of \textit{all} matter in the central regions of bursty gas-rich dwarf galaxies, particularly at high redshifts \citep{Stinson_2009, Maxwell_2012, Teyssier_2013, Governato_2015, Chan_2015}. However, few works have studied in detail the effects of outflows on stellar kinematics, especially at late cosmic times.

Stellar migration has been studied extensively in other contexts, though mostly in massive ($\sim L_*$), disk-dominated galaxies. Here, non-axisymmetric gas and stellar structures within the disk have been shown to drive radial migration without appreciably heating the stellar population \citep{2002MNRAS.336..785S, 2008ApJ...684L..79R, 2013A&A...553A.102D, 2015arXiv151106369L}. However, fewer works have investigated radial migration in low-mass galaxies. Because low-mass galaxies generally lack strong disks, bars, or spiral arms, radial migration often is assumed to be less important in this regime. Some previous studies of low-mass galaxies have found that their stellar orbits are perturbed primarily by gentle dynamical heating \citep{Stinson_2009, Schroyen_2013, Teyssier_2013}, which becomes important only on cosmological timescales (if at all) and does not appreciably change radial population gradients. In general, most studies of stellar kinematics and radial migration assume that the global galactic potential is smooth and varies gradually over time. However, the many recent studies of dark-matter coring, as cited above, demonstrate that this assumption is not necessarily valid in the low-mass regime. 

In this work, we investigate in detail the effects of stellar feedback-driven gas outflows/inflows on \textit{stellar} kinematics in isolated low-mass galaxies using cosmological zoom-in hydrodynamic simulations. We examine how feedback drives rapid radial migration of stars over short timescales and demonstrate that the cumulative effects of many starburst episodes drive systematic outward migration over cosmological timescales. We organize our paper as follows. In Section~\ref{sec:methods}, we describe our simulations and the properties of our galaxy sample at $z = 0$. In Section~\ref{sec:results}, we examine in detail one of our galaxies as a case study, highlighting its bursty star formation and the effects of gas outflows/inflows on stellar kinematics, radial migration, and population gradients. In Section~\ref{sec:scaling_w_mass}, we explore this behavior across our entire galaxy sample. In Section~\ref{sec:comparison}, we discuss the relation between stellar migration and dark matter coring and compare with previous theoretical and observational results. Finally, in Section~\ref{sec:summary}, we summarize our results and discuss possible observational tests.

\section{Methods}
\label{sec:methods}

\subsection{FIRE Simulations}
\label{sec:simulations}

We use cosmological zoom-in hydrodynamic simulations of isolated galaxies from the FIRE (Feedback in Realistic Environments) project\footnote{http://fire.northwestern.edu} \citep{Hopkins_2014}. All simulations were run using the \textsc{GIZMO} code \citep{Hopkins_2015} with pressure-entropy based smoothed particle hydrodynamics \citep[P-SPH;][]{2013MNRAS.428.2840H} and an updated version of the PM+TREE gravity solver from GADGET-3 \citep{Springel_2005}. For more details and tests of \textsc{GIZMO}, see \citet{Hopkins_2015}. All of our runs use a $\Lambda$CDM cosmology with $\left(\Omega_{M}, \Omega_{\Lambda}, \Omega_{b}, h \right) = \left(0.272, 0.728, 0.0455, 0.702 \right)$. Because all of our simulations have been presented in previous works, we briefly summarize only the most important features.

Our simulations incorporate gas dynamics and radiative cooling using tabulated cooling rates from \textsc{CLOUDY} \citep{Ferland_2013} across $10 – \selectlanguage{english}10 ^ {10}$ K, which include cooling from atoms, metals (using 11 species), and molecules. We include ionization and heating from a redshift-dependent ultraviolet background computed in \citet{Faucher_Gigu_re_2009} and estimate self-shielding in dense gas via an on-the-fly local Jeans-length approximation.

As the simulation evolves, stars form as individual gas particles turn into star particles if three conditions are met:
\begin{enumerate}
\item The local gas density must be $n > n_{\mathrm{SF}}$, where $n_{\mathrm{SF}} = 100h^2\,\mathrm{cm ^ {-3}}\approx50\,\mathrm{cm ^ {-3}}$.
\item Star-forming gas must be locally self-gravitating.
\item Star-forming gas must be molecular, as determined by the molecular fraction calculated from the local column density and metallicity according to \citet{2011ApJ...729...36K}.
\end{enumerate}
If these conditions are met, star particles form with an instantaneous efficiency of 100\% per local free-fall time, though stellar feedback quickly regulates this efficiency within a gas cloud. Each star particle that forms represents a single stellar population with single age and metallicity, assuming a \citet{Kroupa_2002} initial mass function.

Some previous papers using FIRE simulations cited $n > n_{\mathrm{SF}} = 100\,{\rm cm^{-3}}$, inadvertently omitting the $h^2$ term. We cite the correct value here. Regardless of the exact threshold, we emphasize that the most important criterion in the FIRE model is that star-forming gas must be locally self-gravitating: we have tested thresholds of $n_{\rm SF} = 5 - 500\,{\rm cm^{-3}}$ and find no significant differences in galaxy-wide properties.

As star particles evolve, they deposit energy, momentum, mass, and metals into nearby gas particles. We incorporate a comprehensive set of stellar feedback processes, as detailed in \citet{Hopkins_2014}: radiation pressure from massive stars, local photoionization and photoelectric heating, core-collapse and type Ia supernovae with appropriate momentum and thermal energy injection, and stellar winds. We use energy, momentum, mass, and metal return computed directly from \textsc{STARBURST99} \citep{Leitherer_1999}; we never turn off cooling of supernova-heated gas. This feedback is injected into the $\approx 32$ gas particles nearest to a given star particle, with each gas particle receiving a fraction proportional to $h_j ^ 3$, where $h_j$ is the gas particle's kernel length.

We simulate each galaxy individually using the cosmological ``zoom-in'' technique \citep{1985PhDT.........7P}, following the method outlined in \citet{Onorbe_2013}. We first run several lower-resolution dark-matter-only cosmological simulations at uniform resolution to identify isolated halos of interest. We then trace the particles around each halo at $z = 0$ back to their initial conditions at $z = 100$ and reinitialize this Lagrangian volume at higher resolution with dark matter and gas particles using the \textsc{MUSIC} code \citep{Hahn_2011}. We then run these zoom-in initial conditions to $z = 0$. For further details, see \citet{Hopkins_2014} and \citet{Chan_2015}.

\subsection{Galaxy Sample}
\label{sec:sample}

Table~\ref{tab:parameters} summarizes the properties of our simulated galaxies at $z = 0$.\footnote{Because m11v is undergoing a major merger at $z = 0$, we cite its values and carry out our analysis at $z = 0.2$, before the merger begins, not at $z=0$.}
We study 8 galaxies across $\mstar(z=0) = 2 \times 10^{6} \mstar$ to $5 \times 10^{10} \mstar$.
Our analysis focuses primarily on the 7 low-mass ``dwarf'' galaxies with $\mstar(z = 0) < 5 \times 10^{9} \mstar$.
We include a Milky Way-like galaxy (m12i) as a comparison for the low-mass regime. All of these galaxies are isolated at $z \sim 0$, with no more massive halo within at least $5\,\rthm$, where $\rthm$ is the spherical radius enclosing $200 \times$ the average matter density of the Universe.

All our galaxies have been presented in previous papers. Specifically, \citet{Hopkins_2014} first presented m10, m11, m11v, and m12i, showing that these galaxies reproduce the observationally-inferred $\mstar - \mhalo$ relation at all redshifts where observational constraints are available. \citet{Chan_2015} first presented m10.1, m10.2, m10.6, and m11.2\footnote{\citet{Chan_2015} refer to simulations m10.1, m10.2, m10.6, and m11.2 as m10h1297, m10h1146, m10h573, and m11h383, respectively. For m10.2, we study the halo within the zoom-in region that has the largest \textit{stellar} mass, while they studied the halo with the largest \textit{total} mass.} and showed that the physically-motivated FIRE feedback prescription produces realistic dark matter density profiles for all the galaxies in our sample. Furthermore, \citet{2015arXiv150402097M} studied the enrichment histories of m10, m11, m11v, and m12i and showed that these galaxies reproduce the observed redshift-evolution of the mass-metallicity relation; \citet{Muratov_2015} studied m10, m11, and m12i and showed that mass-loss from feedback-driven gas outflows can explain the suppressed star-formation efficiencies of low-mass galaxies; \citet{Faucher_Giguere_2015} studied m10, m11, and m12i at redshifts $z=2-4$ and found that these galaxies reproduce the the dense HI covering fractions of observed Lyman-break galaxies; and \citet{2015arXiv151003869S} studied the burstiness of the SFHs of m10, m11, and m12i and compared this with observed local galaxies, finding that the simulations reproduce the observed main sequence of star formation but may overpredict the quenched fraction at $\mstar \lesssim 10^{9.5}\,\msun$.

\begin{table*}
\caption{
Parameters of the simulations at $z = 0$.}
\label{tab:parameters}
\begin{tabular}{ p{0.9cm}| p{1.4cm} | p{0.8cm} | p{1.4cm} | p{0.6cm} | p {1cm}| p{0.5cm} | p{0.7cm} | p{1.2cm} |p{0.7cm} | p{0.6cm} | p{0.6cm} | p{0.6cm} | p{0.7cm} | p{0.7cm} }
Name     & $\log(M_{200m})$ $\mathrm{[M_{\odot}]}$ & $R_{200m}$ $[\mathrm{kpc}]$ 
    	 & $\log(\mstar)$ $\mathrm{[M_{\odot}]}$ & $R_{e}$ $[\mathrm{kpc}]$ 
    	 & $R_{\rm 90mass}$ $[\mathrm{kpc}]$ & $f_{\mathrm{gas}}$ & $\alpha$ & $\log(N_{\rm star})$ 
    	 & $t_{\rm dyn}$ $\rm{[Myr]}$ & $m_{\rm dm}$ $\mathrm{[M_{\odot}]}$ & $m_{\rm b}$ $\mathrm{[M_{\odot}]}$  
    	 & $\epsilon_{\rm dm}$ $[\mathrm{pc}]$ & $\epsilon_{\rm star}$ $[\mathrm{pc}]$ & $\epsilon_{\rm gas}$ $[\mathrm{pc}]$\\
\hline
m10 	& 9.92 	& 64 	& 6.35 	& 0.17 	& 1.40	& 0.77	& -1.63	& 4.04	& 78	& 1.3e3 & 2.6e2 & 29	& 7	& 3\\
m10.1 	& 10.16 & 77 	& 7.22 	& 0.65 	& 3.97 	& 0.88 	& -0.65	& 3.98 	& 194 	& 1.0e4 & 2.1e3 & 43 	& 7 & 4\\   
m10.2 	& 9.84 	& 61 	& 8.08  & 0.47 	& 2.62  & 0.41  & -0.76	& 4.81  & 160 	& 1.0e4 & 2.1e3 & 43 	& 7 & 4\\
m10.6 	& 10.60 & 110 	& 8.47 	& 1.94 	& 9.02 	& 0.81 	& -0.43	& 5.23 	& 263 	& 1.0e4 & 2.1e3 & 100 	& 21& 10\\
m11 	& 11.17 & 170 	& 9.32 	& 5.26 	& 15.45 & 0.48 	& -0.37	& 5.47 	& 296 	& 3.5e4 & 7.1e3 & 71 	& 14& 7\\
m11v 	& 11.28 & 150 	& 9.36 	& 4.39 	& 14.05 & 0.46 	& -0.34	& 4.73 	& 282 	& 2.8e5 & 5.7e4 & 142 	& 14& 7\\
m11.2 	& 11.23 & 180 	& 9.59 	& 3.99 	& 14.87 & 0.50 	& -0.36	& 5.43 	& 263 	& 8.3e4 & 1.7e4 & 100 	& 21& 10\\
m12i 	& 12.09 & 340 	& 10.74 & 5.35 	& 12.50 & 0.33 	& -1.32	& 5.99 	& 88 	& 2.8e5 & 5.7e4 & 142 	& 50& 20\\\hline
\hline
               
\end{tabular}
\begin{flushleft}
$R_{200m}$ is the radius at which $\rho \left( < R_{200m} \right) = 200 \rho_{\mathrm{matter}}$, where $\rho \left(< R_{200m} \right)$ is the average matter density over a sphere of radius $R_{200m}$.
$M_{200m}$ and $\mstar$ are the total mass and stellar mass inside $R_{200m}$ and $0.1 \, R_{200m}$, respectively.
$R_{e}$ is the effective radius enclosing 50\% of the stellar light above a surface-brightness threshold of $26\,\mathrm{mag \, arcsec^{-2}}$ in the $r$-band; $R_{\rm 90mass}$ is the radius enclosing 90\% of the stellar mass.
$f_{\mathrm{gas}} = M_{\mathrm{gas}} / \left(\mstar + M_{\mathrm{gas}} \right)$ is the gas fraction inside $0.1 \, R_{200m}.$ 
$\alpha$ is central slope of the dark matter density profile $\left(\rho_{DM}\propto r^{\alpha}\right)$ in the interval $r = (1-2\%)R_{200m}.$
$N_{\rm star}$ is the number of star particles inside $R_{\rm 90mass}$.
$t_{\rm dyn} = \sqrt{3\pi/16G\bar{\rho}}$ is the dynamical time within $R_{\rm 90mass}$, using the average density of all matter, $\bar{\rho}$.
$m_{\rm dm}$ and $m_{\rm b}$ are the average particle masses for dark matter and baryons; $\epsilon_{\rm dm}$, $\epsilon_{\rm star}$, and $\epsilon_{\rm gas}$ are the minimum gravitational softening lengths, in physical units.
\end{flushleft}
\end{table*}

\subsection{Calculating Centers and Sizes of Galaxies}
\label{sec:calculating_size}

We define the center of a galaxy (which also defines the center of its host halo) via an iterative zoom-in method using only star particles. We first calculate the center of (stellar) mass using the entire zoom-in region. We then recalculate this iteratively by reducing the search radius by 50\% at each iteration until we identify the position that encompasses only 32 particles. This iterative zoom-in method ensures that we always center on the density peak closest to the center of mass. We also experimented with centering on dark matter rather than stars, finding only modest $(< 10\%)$ differences in our results for radial migration.

In an effort to both measure the most physically meaningful sizes and compare our simulations to observations, we use two size measurements for each galaxy. First, we compute $R_{\rm 90mass}$, the spherical radius that encloses 90\% of the total $\mstar$. We also compute $R_e$, the effective radius that encloses half of the (observable) light.

We calculated $R_e$ as follows. First, we construct a grid of 6150 $r$-band luminosity functions calculated from the most recent Padova isochrones \citep{Bressan_2012} assuming a \citet{Kroupa_2002} initial mass function. These are logarithmically spaced in stellar age and linearly space in metallicity to span the full range of stellar properties in the simulations. We integrate over each luminosity function to obtain a grid of stellar luminosities per unit initial mass as a function of age and metallicity. We assign each star particle a luminosity by interpolating its age and metallicity from the grid and weighting the result by its initial mass. Finally, we mock observe each galaxy along a given line of sight to calculate the surface brightness in elliptical radius bins that are aligned with the galaxy's major axis. Our measurements of $R_e$ do not attempt to account for scattering or dust attenuation.

We define $R_e$ as the semi-major axis of the ellipse that contains 50\% of the light above an $r$-band surface brightness threshold of $26\,\mathrm{mag\,arcsec^{-2}}$ (similar to the SDSS observations against which we compare). We quantified the line-of-sight scatter in these measurements via mock observations along randomly chosen lines of sight. None of our galaxies with $\mstar < 10 ^ {10} \msun$ have axis ratios $< 0.5$, and the $R_e$ calculated along different lines of sight differ by only $\sim 10\%$. As an additional sanity check, we also experimented with fitting Sersic profiles to our galaxies, finding that they generally yield effective radii within 10\% of the radii calculated above.

\subsection{Effects of Resolution}
\label{sec:resolution_dependence}

To test whether our results depend significantly on resolution, we analyzed two of our galaxies (m11 and m11.2) simulated at 8 times lower resolution in mass. In both simulations, the instantaneous SFR is more bursty in the lower-resolution runs, but the burstiness of the SFR averaged over 100 Myr is unchanged. For both galaxies, we find that the fractional radial migration, population gradients, and size fluctuations agree to within a few percent at both resolutions. We thus conclude that, although star formation in the FIRE simulations depends somewhat on resolution \citep[see][]{Hopkins_2014, 2015arXiv151003869S}, the specific processes that we study---radial migration, size fluctuation, and the effect on population gradients---do not depend significantly on resolution.

\section{Case Study: Evolution of a Single Galaxy}
\label{sec:results}

We present our results in two parts. First, in this section, we examine in detail the SFH, stellar and gas kinematics, size evolution, and population gradients of a single galaxy, m10.6, with $\mstar(z=0) \approx 3 \times 10 ^ 8 \msun$ (see Table~\ref{tab:parameters}). We use this galaxy, which is representative of the mass regime in which radial migration is most significant, as a case study to explore in detail both short-timescale behavior across a single gas inflow- outflow episode and long-term trends across cosmic time. Then, in Section~\ref{sec:scaling_w_mass}, we present results for all 8 galaxies to highlight trends with $\mstar$ from $2\times10 ^ 6$ to $5 \times 10^{10} \msun$.

\subsection{Bursty Star Formation Histories}
\label{sec:bursty}

\begin{figure}
	\includegraphics[width=\columnwidth]{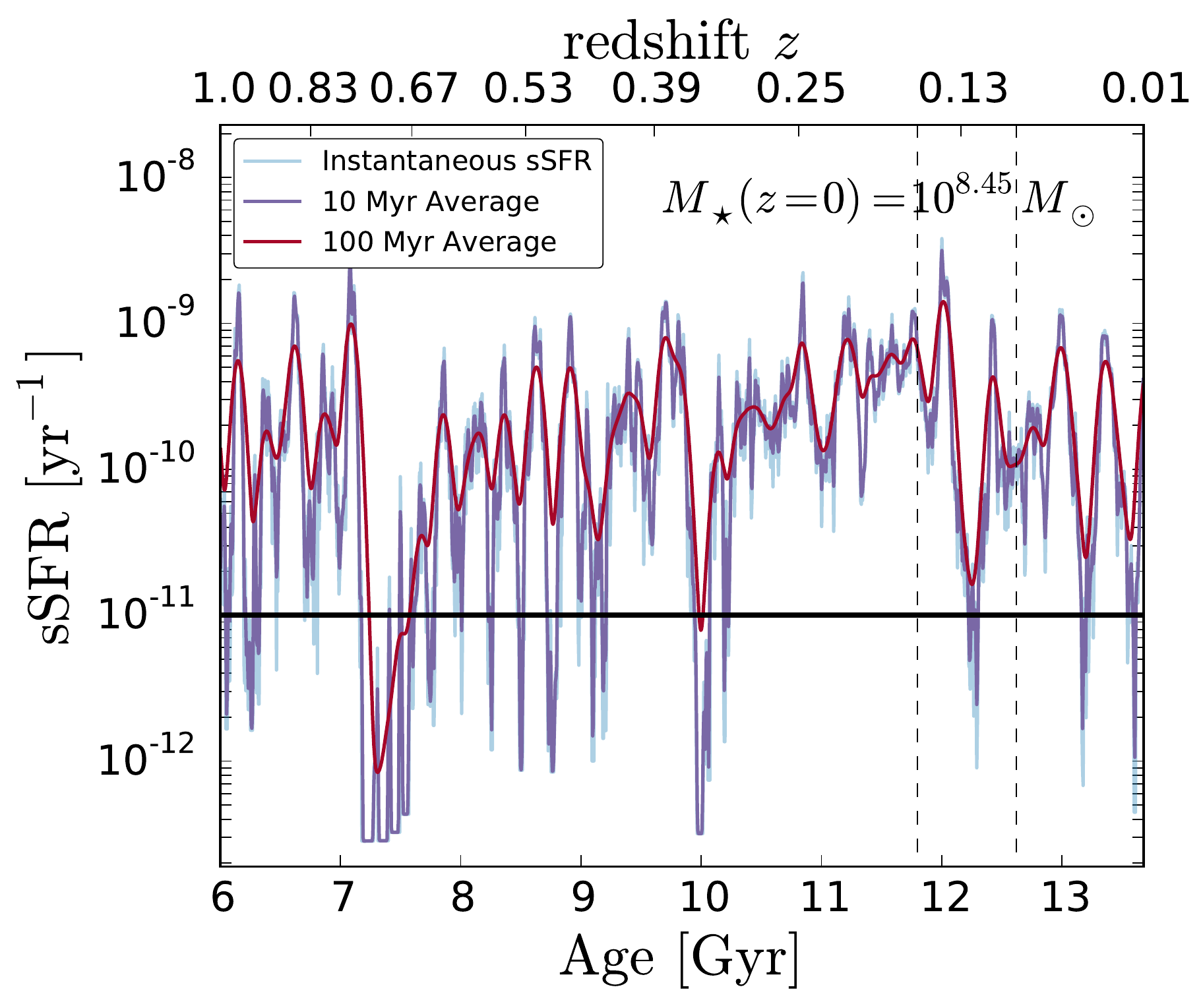}
	\caption{Specific star formation rate (sSFR) in m10.6 from $z = 1$ to $z=0$. Different colored curves show sSFR smoothed over different timescales: 10 (100) Myr approximates the timescale for significant H$\alpha$ (UV) emission. Horizontal black line shows sSFR $= 10^{-11}\,\mathrm{yr^{-1}}$, a common threshold used to define a galaxy as quiescent. The sSFR fluctuates by several orders of magnitude over $< 100$ Myr. Major starburst episodes, which blow much of the cold gas out into the halo and cause the sSFR to temporarily quench (fall below the quiescence threshold), occur several times after $z = 1$. Vertical dashed lines at $\approx 12 \gyr$ ($z \approx 0.13$) enclose a single starburst episode, which we examine in detail below.}
	\label{fig:quenched_percentage}
\end{figure}

The SFHs of our low-mass galaxies are highly stochastic, even at late cosmic times. Figure~\ref{fig:quenched_percentage} shows the specific star formation rate ($\mathrm{sSFR=SFR/M_{\star}}$) of m10.6 from $z = 1$ to $0$. We focus on late-time evolution to highlight behavior at redshifts where such low-mass galaxies currently are observable.

Bursty star formation (even at late times) is typical of our simulated galaxies near this mass \citep{2015arXiv151003869S}. This burstiness is driven by stellar feedback, which drives gas outflows and regulates the central gas density on short timescales. Following a burst of star formation, young stars and supernovae inject energy and momentum into the ISM to drive gas out into the halo, temporarily disrupting star formation. However, because most outflowing gas is not accelerated past the escape velocity at late times \citep{Muratov_2015}, it rapidly cools and re-accretes into the center of the galaxy, where the SFR rises again. Figure~\ref{fig:quenched_percentage} illustrates this cycle in m10.6; the SFR falls each time the sSFR approaches $10^{-9}\,\mathrm{yr^{-1}}$. These fluctuations in SFR are semi-periodic, with a typical spacing of a few 100 Myr, which is comparable to the galaxy's dynamical time (Table~\ref{tab:parameters}).

As Figure ~\ref{fig:quenched_percentage} shows, m10.6 goes through $\sim 10$ periods of strongly increased, then decreased, sSFR between $z = 1$ and $0$. Even averaged over 10 and $100 \myr$, the typical timescales over which observable $\halpha$ and UV emission are enhanced, these reduced sSFRs briefly fall below the typical threshold of sSFR $ < 10 ^ {-11} \yrinv$, where galaxies are classified as quiescent.

For the rest of this section, we focus in particular on evolution within the 400-Myr burst episode at $z \approx 0.13$, as shown by the vertical dashed lines in Figure~\ref{fig:quenched_percentage}. The trends that we present are typical during such burst episodes for our galaxies near this mass.

\subsection{Kinematics and Morphology of Gas and Stars}
\label{sec:kinematics}

\begin{figure*}
	\begin{center}
	\includegraphics[width=0.8\textwidth]{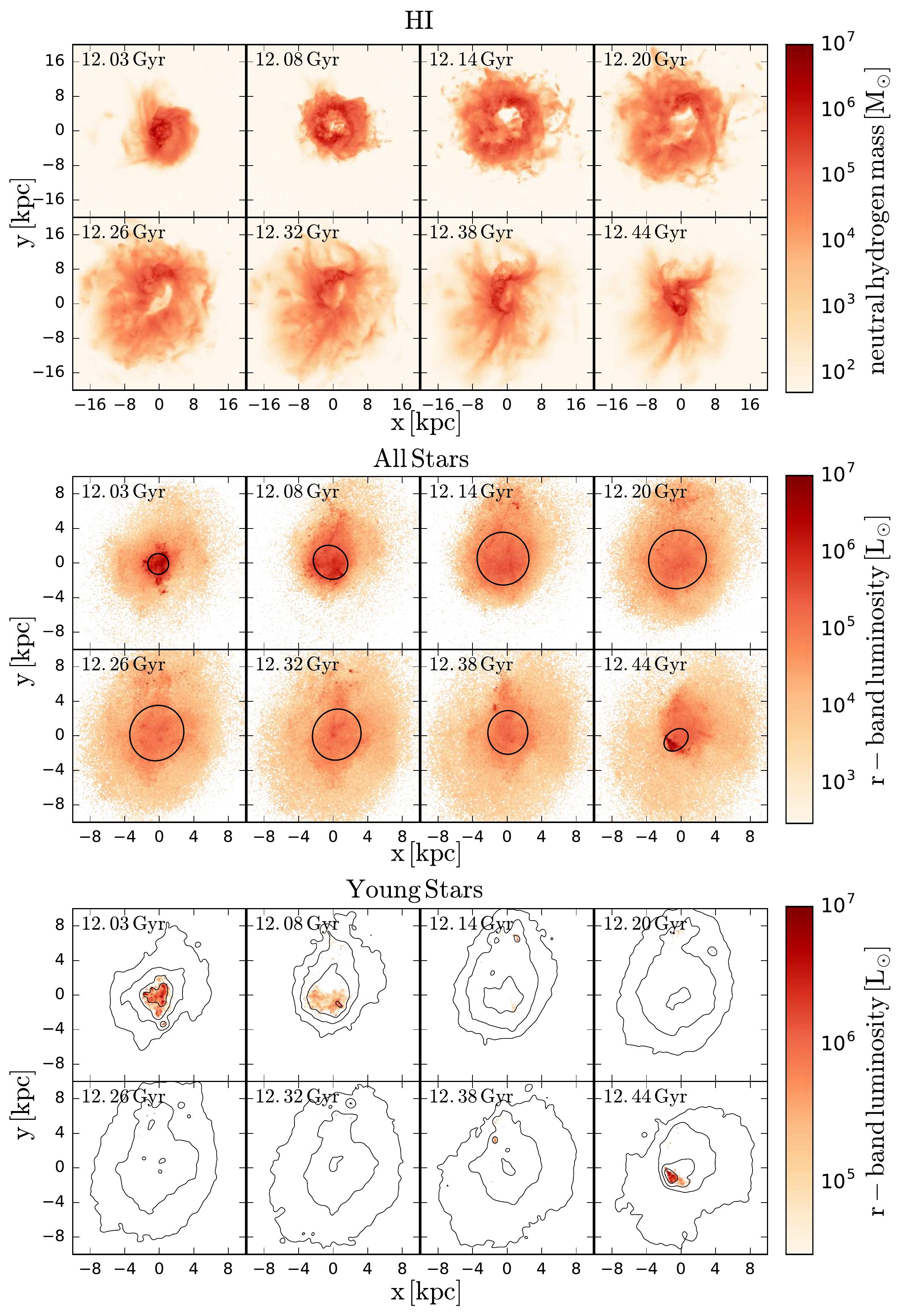}
	\caption{Evolution of neutral gas (top), all stars (middle), and young stars (bottom) in m10.6 across a single 400-Myr starburst episode (vertical lines in Figure~\ref{fig:quenched_percentage}). Panels show snapshots spaced by $\approx 60 \myr$; sSFR in each snapshot is shown by blue points in Figure~\ref{fig:sfr_vs_vr}. Note the difference in spatial scale between the top and bottom two panels. 
\textbf{Top}: Distribution of neutral atomic hydrogen. The SFR is highest in the first panel, when the gas is densest. The SFR reaches its lowest value at $t=12.26 \gyr$, when nearly all the gas is blown beyond the stellar distribution. Once the gas cools back into the center, the SFR rises quickly again, nearly returning to its pre-outflow value by $t=12.44 \gyr$.
\textbf{Middle}: Projected distribution of stellar light (in the $r$-band). Black ellipses enclose show $R_e$, the radius enclosing 50\% of the light. The stellar distribution expands and contracts similar to the gas, with $R_e$ varying by a factor of $\approx 2.5$, across this timescale.
\textbf{Bottom}: Projected distribution of stellar light for stars younger than 50 Myr (smaller than the time spacing between panels). Black contours show logarithmically spaced luminosity of all stars (from middle panel). Some stars continue to form during the beginning of the outflow phase, but almost no stars form after this, when gas is diffuse, ionized, and not self-bound.}
	\label{fig:gas_star_distribution}
	\end{center}
\end{figure*}

We first examine the influence of stellar feedback and bursty star formation on the kinematics and morphology of cold gas, and its consequent influence on the kinematics and morphology of stars, during the single starburst/quenching episode highlighted in Figure~\ref{fig:quenched_percentage}.

Figure~\ref{fig:gas_star_distribution} shows the surface density of neutral atomic hydrogen (top), all stars (middle), and young stars (age $< 50 \myr$) across 8 snapshots spaced by $\approx 60 \myr$. (Note the change in spatial scale between the top and middle/bottom panels.) In the first snapshot, the SFR is high and concentrated within the inner $\approx 2 \kpc$. Stellar feedback then creates a nearly symmetric galactic wind, which drives the gas almost entirely beyond the stellar distribution: at the extremum of the outflow, the central $\approx 5 \kpc$ are almost entirely devoid of cold gas. However, because these winds are launched below the escape velocity and the cooling time of the gas remains short \citep{Muratov_2015}, the gas soon reaccretes back into the galaxy, and another round of star formation begins.

This expansion and contraction is not limited to the gas. While gas is the only species that couples directly to stellar feedback, this low-mass galaxy is sufficiently gas-rich (Table~\ref{tab:parameters}) that the displacement of outflowing gas leads to significant fluctuations in the global galactic potential. Thus, the kinematics of the (effectively) collisionless stars and dark matter also change as a result of feedback. Figure~\ref{fig:gas_star_distribution} (middle) shows the $r$-band luminosity surface density for all stars. Critically, the stellar distribution expands and contracts in the same way and over the same timescale as the gas. This expansion is present both at the center, where the surface density drops significantly, and in the outskirts, where the stars expand to even larger radii. The black ellipses show the 2-D fit to $R_{e}$ at each snapshot, highlighting that within just $\approx 200 \myr$, $R_{e}$ increases by more than a factor 2. Similar fluctuations occur for dark matter \citep{Chan_2015}.

Figure~\ref{fig:gas_star_distribution} (bottom) shows the $r$-band luminosity surface density of only stars younger than 50 Myr, slightly less than the time spacing between these snapshots. This highlights the effect of stellar feedback on star-forming regions. Initally, star formation occurs primarily just within the inner $\approx 2 \kpc$. During the outflow period, the SFR decreases, but it does not terminate immediately: some young stars are formed in gas that is already outflowing.

\begin{figure}
	\includegraphics[width=\columnwidth]{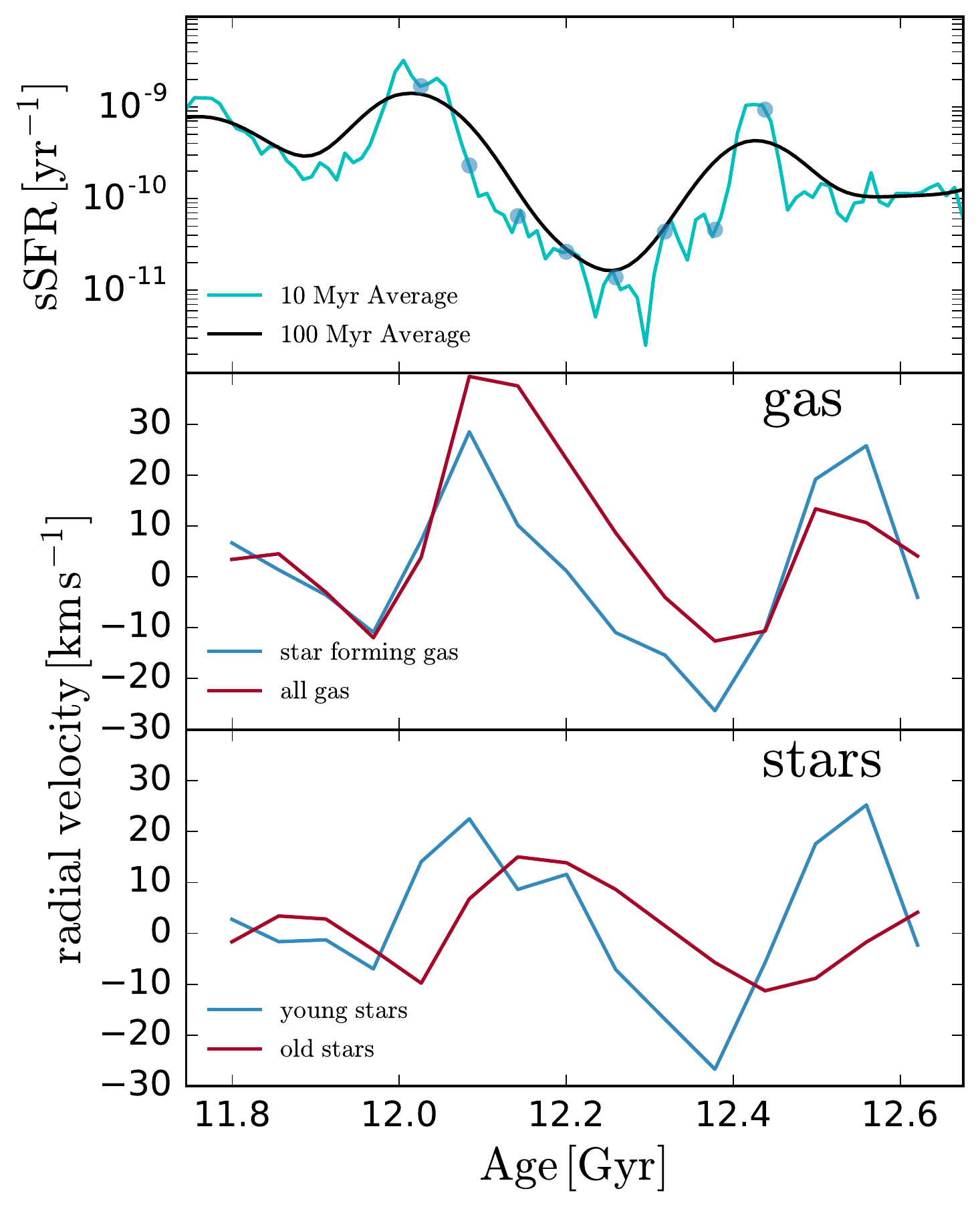}
	\caption{Short-timescale effects of stellar feedback on the kinematics of gas and stars across a single starburst episode (vertical lines in Figure~\ref{fig:quenched_percentage}) in m10.6. 
\textbf{Top}: Specific star formation rate (sSFR) averaged over timescales of 10 (cyan) and 100 (black) Myr. Blue points show time-spacing of the 8 snapshots shown in Figure~\ref{fig:gas_star_distribution}.
\textbf{Middle}: Average radial velocity of gas within 25 kpc, for star-forming (blue; ee Section ~\ref{sec:simulations}) and all (red) gas. The radial velocity of gas closely traces the evolution of the SFR.
\textbf{Bottom}: Average radial velocity of stars, both young (age $< 10 \myr$, blue) and old (age $> 2 \gyr$, red). The radial velocity of stars closely traces that of gas. Young stars, inheriting the kinematics of star-forming gas, showing stronger kinematic fluctuations than old stars, but even old stars respond to the rapidly varying potential during the gas outflow-inflow cycle.}
	\label{fig:sfr_vs_vr}
\end{figure}

Figure~\ref{fig:sfr_vs_vr} quantifies the kinematics of both gas and stars by showing their average radial velocity ($v_{\rm r}=\frac{\mathbf{v}\cdot\mathbf{r}}{\left|\mathbf{r}\right|}$) as a function of time over 1 Gyr spanning the same 400-Myr period as Figure~\ref{fig:gas_star_distribution}. In all the following plots, averages are weighted by particle mass. The top panel shows the sSFR to highlight its correlation with gas and stellar kinematics. The middle panel shows the radial velocity of all gas (red) and instantaneously star-forming gas (blue) within $25 \kpc$. Finally, the bottom panel shows the radial velocity of extremely young stars (age $< 10 \myr$, blue) and old stars (age $> 2 \gyr$, red). Overall, the radial velocities of both gas and stars correlate with the sSFR remarkably closely.

The sSFR reaches its peak at $t \approx 12 \gyr$, just after gas has (re)accreted into the center. Then, just $\sim 100$ Myr after peak sSFR, stellar feedback has driven gas to its maximum outward velocity of $30 - 40 \kms$ (on average), at which point sSFR declines most rapidly. As the gas approaches turn-around ($v_{\rm r} = 0$) and is most rarefied, the sSFR reaches its minimum. Finally, as the gas starts to cool back into the galaxy, the sSFR starts to increase, and the cycle continues. During this outflow period, less than 2\% of all gas is accelerated beyond the escape velocity ($v_{{\rm esc}} \approx 160 \kms$ in the center of this galaxy), and none of the galaxies in our sample experience significant mass loss after $z=1$ \citep{Muratov_2015}.

Figure~\ref{fig:sfr_vs_vr} (middle) shows that even star-forming gas clouds participate in this inflow-outflow cycle. That is, some outflowing/infalling gas remains in dense ($n_{\rm H}>50\,\mathrm{cm^{-3}}$), cool ($T\lesssim10^{3}\mathrm{\,K}$), self-bound clumps with ongoing star formation. Figure~\ref{fig:sfr_vs_vr} (bottom) shows that the radial velocity of young stars closely traces that of star-forming gas. Thus, \textit{young stars inherit the complex inflow-outflow kinematics of the gas clouds in which they form}. Furthermore, the majority of all stars in our simulated galaxies with $\mstar \lesssim 10 ^ {10} \msun$ form during burst episodes \citep{2015arXiv151003869S}; thus, a majority of stars can experience non-trivial radial migration within their first $\sim 100 \myr$.

Figure~\ref{fig:sfr_vs_vr} (bottom) shows a systematic time offset between the radial velocities of young and old stars. Young stars (almost by definition) are strongly coupled to the gas clouds in which they form. Stars with intermediate ages of a few 100 Myr (not shown) start to separate kinematically from their progenitor gas clouds. Importantly, \textit{old stars still fluctuate kinematically}, though their fluctuations are not as strong as those of young stars or gas. The delayed response of old stars demonstrates that they do not respond directly to the kinematics of gas, but instead, respond to the change in the galactic potential when feedback drives significant gas mass beyond the stellar component.

It is important to note that although gas and stars move in and out \textit{on average} during the burst cycle, some particles on radial orbits do move inward during outflow periods and vice-versa. At the peak of the outflow shown in figure 3, approximately 87\% of gas and 73\% of stars have positive radial velocities, with $\left(\sigma_{r,\,{\rm gas}},\,\sigma_{r,\,{\rm star}}\right)\approx\left(39,\,26\right)\,{\rm km\,s^{-1}}.$ We have verified that the average velocities shown in Figure ~\ref{fig:sfr_vs_vr} are not driven primarily by high-velocity outliers; the mean and median radial velocities typically agree within a few percent.

All together, we summarize the following two-stage physical picture for the cause of strong radial migration of stars in low-mass galaxies, where stellar feedback drives strong outflow-inflow cycles of gas:
\begin{itemize}
\item (1) Outflowing/infalling gas clouds can remain star-forming, producing young stars that inherit the radial kinematics of their progenitor gas clouds and thus migrate significantly within their first $\sim 100 \myr$.
\item (2) Gas outflows/inflows in low-mass galaxies with high gas fractions drive strong fluctuations in the galaxies' overall potentials, which in turn drive strong kinematic fluctuations in stars of all ages.
\end{itemize}

These processes are fundamentally different from the processes thought to cause radial migration in massive disk galaxies, where stellar orbits are scattered via interactions with massive non-axisymmetric structures \citep[for example,][]{1977A&A....60..263W, 1988MNRAS.230..597B}, or are stirred by spiral waves and/or bars \citep[for example,][]{2002MNRAS.336..785S, Minchev_2011}. In our low-mass galaxies, the entire stellar distribution ``breathes'' (expands and contracts) due to \textit{global} processes resulting from stellar feedback and bursty star formation. This, together with radial kinematics inherited from outflowing/infalling gas, is the primary driver of radial migration in our simulations. Unlike the well-studied spiral wave-driven migration processes that are important in cool disks, the processes driving migration in our simulations also kinematically heat the stellar population and place some stars on highly radial orbits.

\subsection{Radial Migration of Stars}

We next explore how the above complex radial kinematics drive radial migration of stars over both short and long timescales.

\subsubsection{Measuring Radial Migration}
\label{sec:measuring_migration}

We define radial migration as any change in the galactocentric radius of a star particle since its formation. To measure radial migration since formation, we identify the first simulation snapshot that occurs immediately after the formation time of each star particle, at which point we record the particle's radius with respect to the center of the main galaxy. We then compute the radial migration as $\Delta r = r_{\mathrm{current}} - r_{\mathrm{form}}$, the difference between the particle's current radius in a given snapshot and its radius in the snapshot immediately following its formation. 

Our typical time spacing between snapshots is $\approx 50 \myr$. This finite time separation  means that the radius that we measure after formation does not exactly equal the radius at which the particle formed, so our values are subject to some extremely short-timescale migration. We tested this by examining only star particles that formed $< 10 \myr$ before the snapshot at which we measure $r_{\rm form}$. We find no systematic difference in any of our results when examining only these particles.

To avoid ambiguity from stars that formed in different (satellite) galaxies that then merged with the main galaxy, we ignore star particles that formed at $r > 2\,R_{\rm 90mass}(z)$, where $R_{\rm 90mass}(z)$ is the radius enclosing 90\% of $\mstar$ of the main galaxy at a given $z$. For several of our galaxies, this means that we exclude $\approx 50\%$ of the stars that formed before $z \sim 2$ because of the higher merger rates at early times.

Throughout, we present both $\langle \Delta r \rangle$, the average of the \textit{net} (vector) radial migration distance across all particles, and $\langle \left| \Delta r \right| \rangle$, the average of the \textit{absolute} radial migration distance. The former measures systematic inward/outward migration, while the latter measures the scatter induced by simultaneous inward and outward migration.

\subsubsection{Radial Migration over Short Timescales}
\label{sec:migration_short}

\begin{figure}
	\includegraphics[width=\columnwidth]{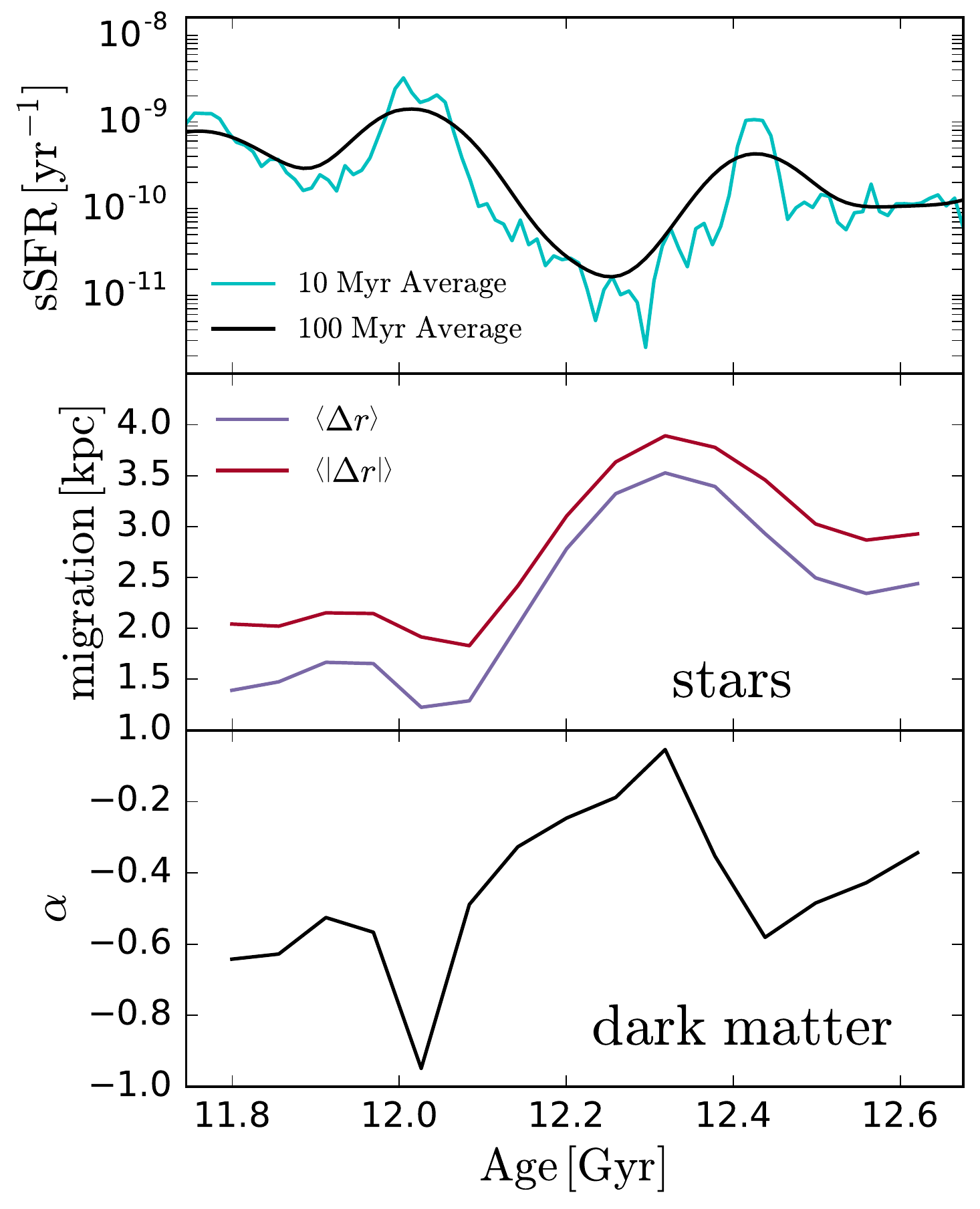}
	\caption{Changes in the distribution of stars and dark matter in m10.6 across the same starburst episode
shown in Figure~\ref{fig:sfr_vs_vr}.
\textbf{Top}: Specific star formation rate (sSFR) averaged over timescales of 10 (cyan) and 100 (black) Myr.
\textbf{Middle}: Mean radial migration of stars relative to their formation radius. 
$\langle \Delta r \rangle$ (purple) shows the net radial migration, while 
$\langle \left| \Delta r \right| \rangle$ shows the absolute radial distance. Migration 
correlates strongly and inversely with sSFR. During this outflow episode, the half-mass 
radius increases from 2.5 to $> 5$ kpc within $\approx 200$ Myr.
\textbf{Bottom}: Central slope $\alpha$ of the dark matter density profile 
$\left(\rho_{DM}\propto r^{\alpha}\right).$ $\alpha$ correlates with mean stellar migration,
since star and dark matter particles feel the same time-varying gravitational potential.}
	\label{fig:sfr_vs_migration}
\end{figure}

Figure~\ref{fig:sfr_vs_migration} compares the time-evolution of the sSFR, the mean stellar radial migration, and the slope of the central dark-matter density profile over the same starburst episode shown in Figure~\ref{fig:sfr_vs_vr}. The middle panel shows the average migration distance for all stars. Like radial velocity, radial migration distance is closely related to the sSFR: stars migrate outwards when the sSFR falls during periods of net outflow and migrate inward once gas falls back into the galactic center and the sSFR rises. This has a dramatic effect on the radial distribution of stars: during the main outflow episode, the overall stellar population migrates coherently $\approx 2.5 \kpc$ outward in less than $300 \myr$. Note that $\langle \Delta r \rangle$ is always positive because we measure radial migration \textit{since formation}, and as we will show in Section ~\ref{sec:migration_long}, stars experience coherent and lasting outward migration over long timescales.

The bottom panel shows the time-evolution of the central slope $\alpha$ of the dark matter density profile $\left(\rho_{DM}\propto r^{\alpha}\right).$ We define $\alpha$ as the power law which best fits $\rho_{DM}$ in the $r = \left(1-2\%\right)R_{200m}$ interval; see \citet{Chan_2015} for further discussion of $\alpha.$ Here, $\alpha\sim0$ represents a flat central density profile (a ``core''), while $\alpha\lesssim-1$ represents a steep NFW-like profile (a ``cusp.'') Because dark matter particles migrate in and out on the same timescale as stars, $\alpha$ evolves similarly to the mean stellar migration and sSFR, changing from a cuspy profile at peak sSFR to a core at peak outflow.

\subsubsection{Radial Migration over Long Timescales}
\label{sec:migration_long}

\begin{figure}
\includegraphics[width=\columnwidth]{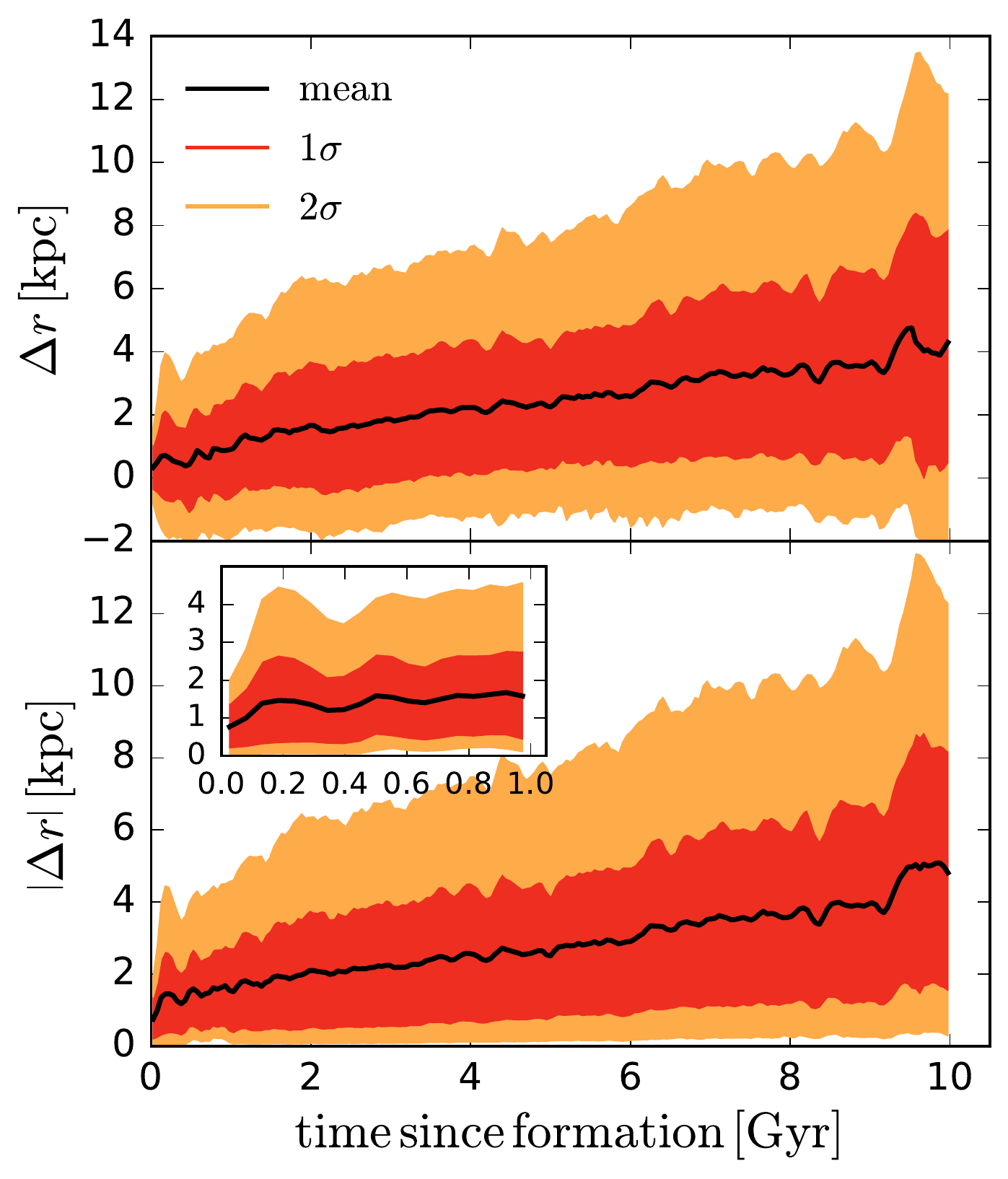}
\caption{Distribution of radial migration distances of stars since their formation as a function of their stellar age in m10.6.
\textbf{Top}: $\Delta\,r$, the difference between a star particle's radius when it is at a given age and its radius when it formed.
Positive (negative) values correspond to stars that have migrated outward (inward) since formation.
\textbf{Bottom}: Same, but for $\left| \Delta\,r \right|$, the absolute radial migration distance.
Inset shows stars younger than 1 Gyr, highlighting short timescales.
After $\lesssim 200 \myr$, stars move an average absolute radial distance of 1 kpc.
As the top panel shows, the average coherent (net) migration is weak over this timescale, but over longer timescales ($\gtrsim 1 \gyr$), stars show strong systematic outward migration via repeated inflow-outflow episodes that drive stellar orbits to larger radii.
Thus, older stars have experienced systematically stronger outward radial migration.}
\label{fig:all_migration}
\end{figure}

We next examine stellar radial migration in m10.6 over long (cosmological) timescales. Figure~\ref{fig:all_migration} shows the distribution of radial migration distances since formation, $\langle \Delta r \rangle$ and $\langle \left| \Delta r \right| \rangle$, as a function of stellar age. Here we stack results across all snapshots, measuring each particle's migration distance in each snapshot and binning the simulation in stellar age. Thus, young ages include a combination of stars measured at $z \sim 0$ and high $z$, while old ages necessarily come from stars measured only at $z \sim 0$.

First, examining $\langle \left| \Delta r \right| \rangle$ in Figure~\ref{fig:all_migration} (bottom), stars undergo a significant fraction of their absolute migration within $\lesssim 200 \myr$, consistent with Figure~\ref{fig:sfr_vs_migration}. The inset zooms in on young ages, showing an early peak of $\approx 1 \kpc$ just $200 \myr$ after formation, the typical duration of outflow episodes. After this initial peak, the absolute migration continues to increase systematically with stellar age. Most importantly, \textit{the oldest stars, which have undergone the largest number of burst episodes, have migrated the most}.

Second, examining $\langle \Delta r \rangle$ (top), young stars ($\lesssim 200 \myr$) show a strong scatter in their net radial migration distance, consistent with the bottom panel, but they do not show systematic net inward/outward migration over such short times. That is, \textit{short-timescale migration is limited to temporary outward/inward burst cycles}. However, over sufficiently long ($\gtrsim 1 \gyr$) timescales, stars show systematic and coherent outward migration that continues to increase with stellar age. This long-term behavior is caused by the repeated semi-periodic oscillations of the potential that gradually heat the orbits of stars over time \citep[for example,][]{Pontzen_2012}, such that the oldest stars, which have undergone the most outflow episodes, have migrated outward the most. Indeed, the oldest stars ($> 8 \gyr$) have migrated outward an average of $\sim 4 \kpc$, or $2\,R_e(z= 0)$, from the radius where they formed.

\subsection{Impact of Radial Migration on Populations Gradients}
\label{sec:population_gradients}

Our simulated galaxies develop significant radial population gradients by $z=0$, with the youngest, most metal-rich stars concentrated near the galactic center and the outskirts dominated by old, metal-poor stars. These gradients are similar to those observed in low-mass galaxies in the local Universe \citep[for example,][]{Mateo_1998, Kirby_2012, Vargas_2015}. Galactic archeology studies, which attempt to infer the formation-history of a galaxy base on its properties at $z \sim 0$, commonly assume that one can translate population gradients observed at $z \sim 0$ to radial star formation and/or chemical enrichment histories. For example, \citet{Zhang_2012} concluded that the star-forming regions of most nearby low-mass have been shrinking because the observed average stellar age increases with radius.

This common assumption is valid if the radial distribution of stars at $z = 0$ accurately reflects the distribution of stars at the time of their formation. However, dynamical heating can can mix different stellar populations and thus alter radial population gradients, and radial migration can preferentially affect old stars \citep{Governato_2015, 2015AAS...22512901B, 2015arXiv151203538G}. The significant radial migration which we present above suggests that, in low-mass galaxies, SFHs that are calculated from population gradients at $z = 0$ may be contaminated significantly by migration. We now investigate this possibility in m10.6.

\begin{figure*}
	\includegraphics[width=\textwidth]{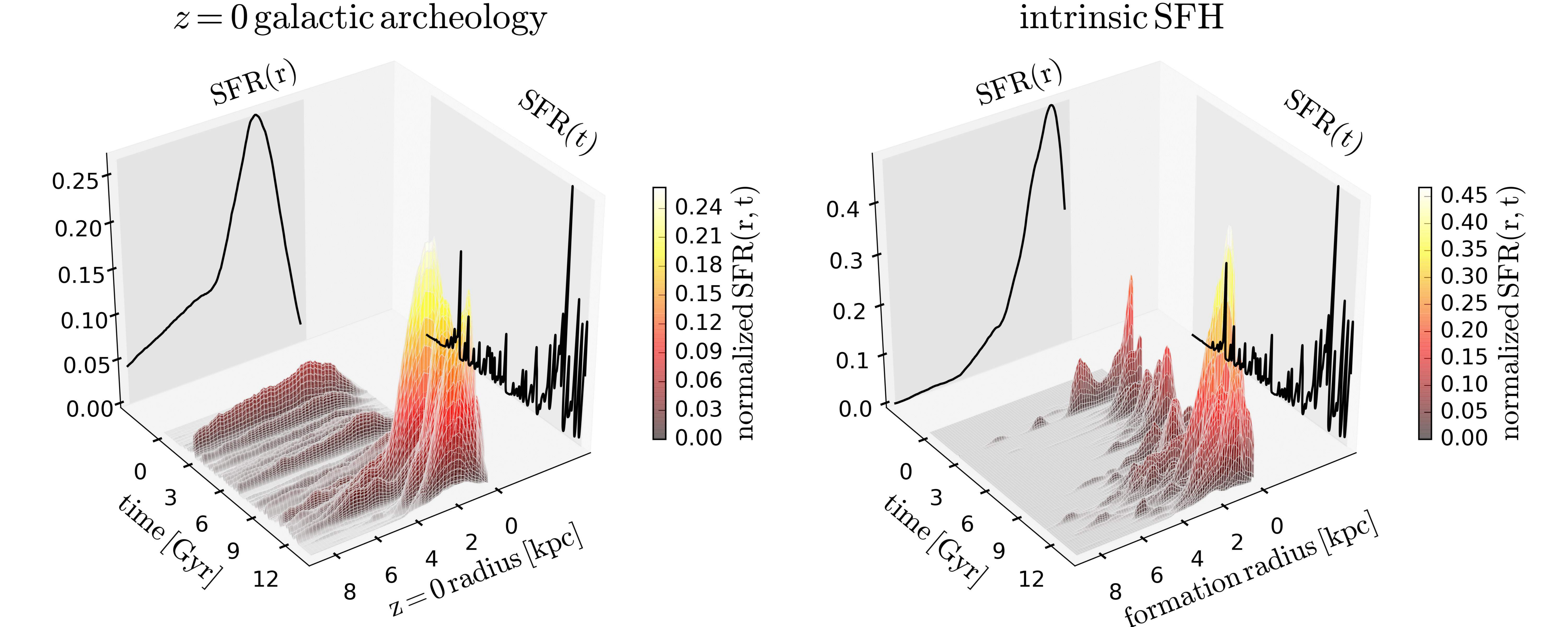}
	\caption{Star formation history (SFH) of m10.6 as a joint function of both time and galactocentric radius. In both versions, the right projection shows the normalized star formation rate (SFR) versus time, while the left projection shows the integral of SFR (mass density) versus radius.
\textbf{Left}: SFH as calculated from the radial distribution of stars at $z = 0$: we bin star particles by their radius at $z = 0$ and compute the SFH in each radius bin, similar to observational approaches for nearby galaxies.
\textbf{Right}: Intrinsic SFH of stars at formation: we bin star particles by their radius at formation (rather than at $z = 0$) and show the \textit{true} radial SFH, without the post-formation effects of radial migration. The differences between these panels, especially for the oldest stars, demonstrate that radial migration can significantly bias the inferred radial SFHs of low-mass galaxies, a critical systematic for galactic-archaeology studies.}
	\label{fig:2dSFH}
\end{figure*}

Figure~\ref{fig:2dSFH} shows the SFH of m10.6 as a joint function of both lookback time and radius. In both versions, the right projection shows SFR versus time, while the left projection shows the azimuthally integrated SFR (density) versus radius.

The left panel shows the SFH if we use the radial distribution of stellar ages at $z = 0$ to infer the SFH, as observational studies do. The radial SFH calculated at $z = 0$ appears to show that early star formation occurred at all radii and in fact was distributed almost uniformly with radius out to $9 \kpc$; by contrast, the youngest stars appear to have formed within the central regions. At face value, this appears to support an ``outside-in'' quenching scenario \citep[for example,][]{Hidalgo_2003, Zhang_2012}, wherein star formation becomes increasingly centrally concentrated as a galaxy evolves.

However, the right panel of Figure~\ref{fig:2dSFH} shows the true underlying SFH, which we measure by tracing the stars back to the radii at which they formed. Thus, this SFH removes any effects of post-formation migration. In reality, star formation has been concentrated in the core since early times: the oldest stars preferentially formed at smaller radii, while the youngest stars formed at systematically larger radii. Thus, the significant effects of radial migration, over both short and long timescales, \textit{qualitatively} have changed the inferred radial SFH in m10.6. This represents a critical systematic in any galactic-archeaology approaches.

\begin{figure}
	\includegraphics[width=\columnwidth]{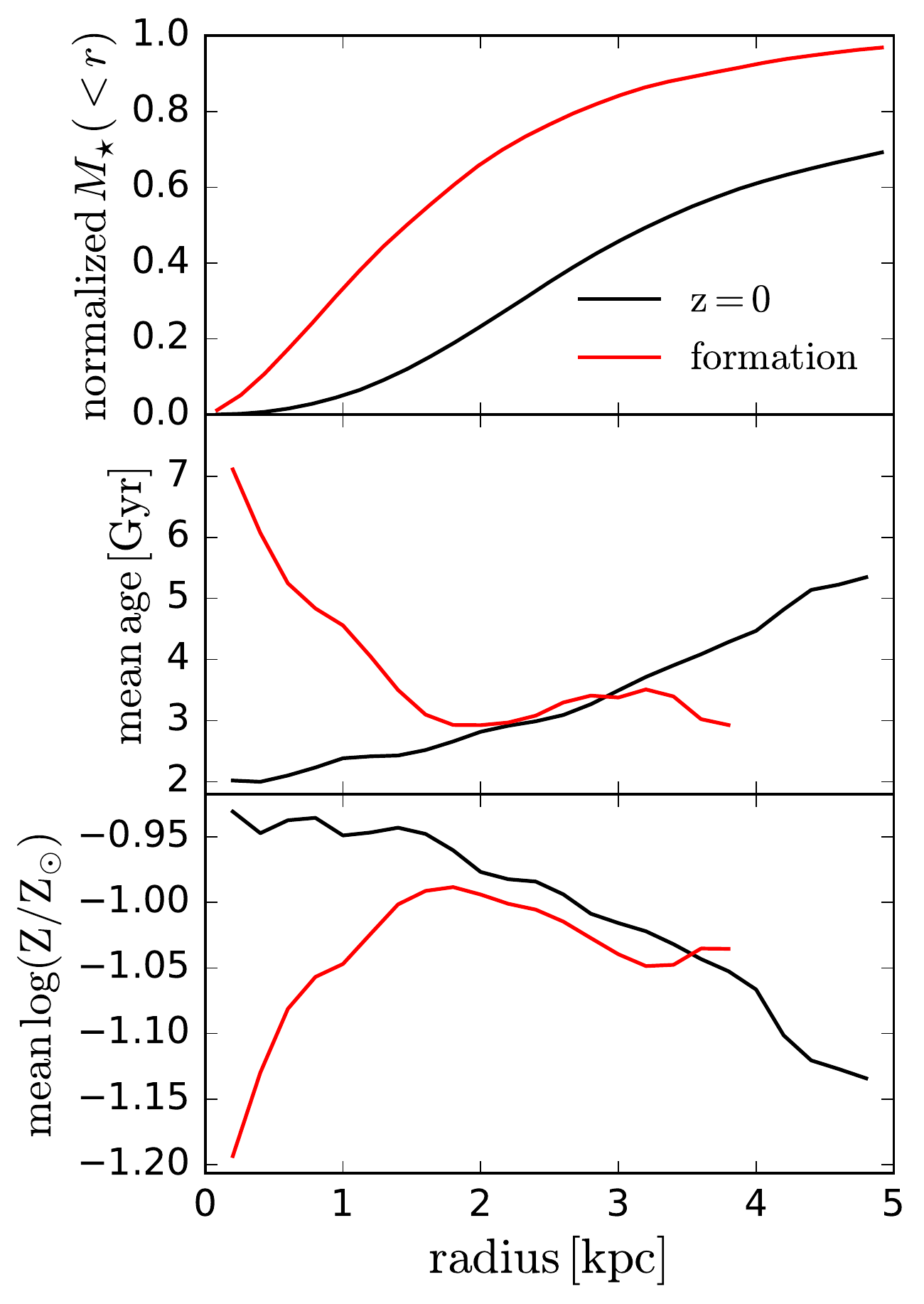}
	\caption{Radial distribution of stellar properties in m10.6. Black curves show stars measured at their radius at $z = 0$, while red curves show stars measured at their formation radius.
\textbf{Top}: Cumulative stellar mass within the given radius. Black curve shows the mass profile at $z = 0$, while red curve shows the mass profile if stars stayed at their formation radius. While $\sim 90\%$ of all stars formed within $\sim 3.5$ kpc, $R_{\rm 90mass}(z=0) > 8 \kpc$, as caused by significant outward migration.
\textbf{Middle}: Mean stellar age as a function of radius at $z = 0$ (black) and at formation (red). Age-dependent radial migration has inverted the true age gradient: while the age gradient measured at $z = 0$ naively implies that younger stars formed preferentially at smaller radii, in reality, the oldest stars formed preferentially at small radii but experienced stronger outward migration. (See also Figure~\ref{fig:2dSFH}.)
\textbf{Bottom}: Mean total metallicity as a function of radius at $z = 0$ (black) and at formation (red).
As with the age gradient, the stronger outward migration of older stars has inverted the metallicity gradient. In reality, older metal-poor stars formed only at small radii but then migrated outward, while younger metal-rich stars formed at all radii and experienced less migration.}
	\label{fig:now_formation}
\end{figure}

Figure~\ref{fig:now_formation} demonstrates these effects more quantitatively. The black curves show stars measured at their radius at $z = 0$, while the red curves show stars measured at their formation radius. The top panel shows the cumulative stellar mass profile: the black curve shows the mass profile at $z = 0$, while the red curve shows the mass profile if stars stayed at their formation radius. Even though all stars form in a centrally concentrated manner, with 90\% having formed within $3.5 \kpc$, the long-term effects of stellar feedback have led to significant outward migration, with $R_{\rm 90mass}(z = 0) > 8 \kpc$.

The middle panel shows the average age of the stellar population as a function of both radius measurements. The red curve highlights that the oldest stars formed only at small radii, while the youngest stars formed across a broad range of radii. Thus, the intrinsic age profile is systematically older in the core and younger at larger radii. However, as Figure~\ref{fig:all_migration} shows, because the oldest stars experienced more starburst cycles, they suffer from stronger outward migration, which has driven them to larger radii. By contrast, the younger stars experienced weaker outward migration, so their current radius more accurately reflects their formation radius. Thus, age-dependent radial migration has inverted the true age gradient.

The bottom panel shows similar behavior for the metallicity gradient. For each star particle, we measure the total metallicity (by mass fraction) and scale this to the solar value assuming $Z_{\odot} = 0.02$. The oldest stars, which were the most metal-poor, formed at small radii but experienced more outward radial migration, while the younger more metal-rich stars formed at larger radii, on average, but experienced less outward migration. Again, radial migration has inverted the true metallicity gradient.

Thus, we conclude that \textit{stellar radial migration, induced by feedback-driven outflows, not only can dilute intrinsic population radial gradients, but also can invert them entirely}.

\section{Dependence on Galaxy Mass}
\label{sec:scaling_w_mass}

Having explored stellar kinematics, radial migration, and population gradients in detail for a single galaxy, m10.6, we now explore these trends for all 8 galaxies in our sample, which span $\mstar(z = 0) = 10 ^ {6.3 - 10.7} \msun$ (or halo $M_{\rm 200m} = 10 ^ {10 - 12} \msun$).

\begin{figure*}
	\includegraphics[width=\textwidth]{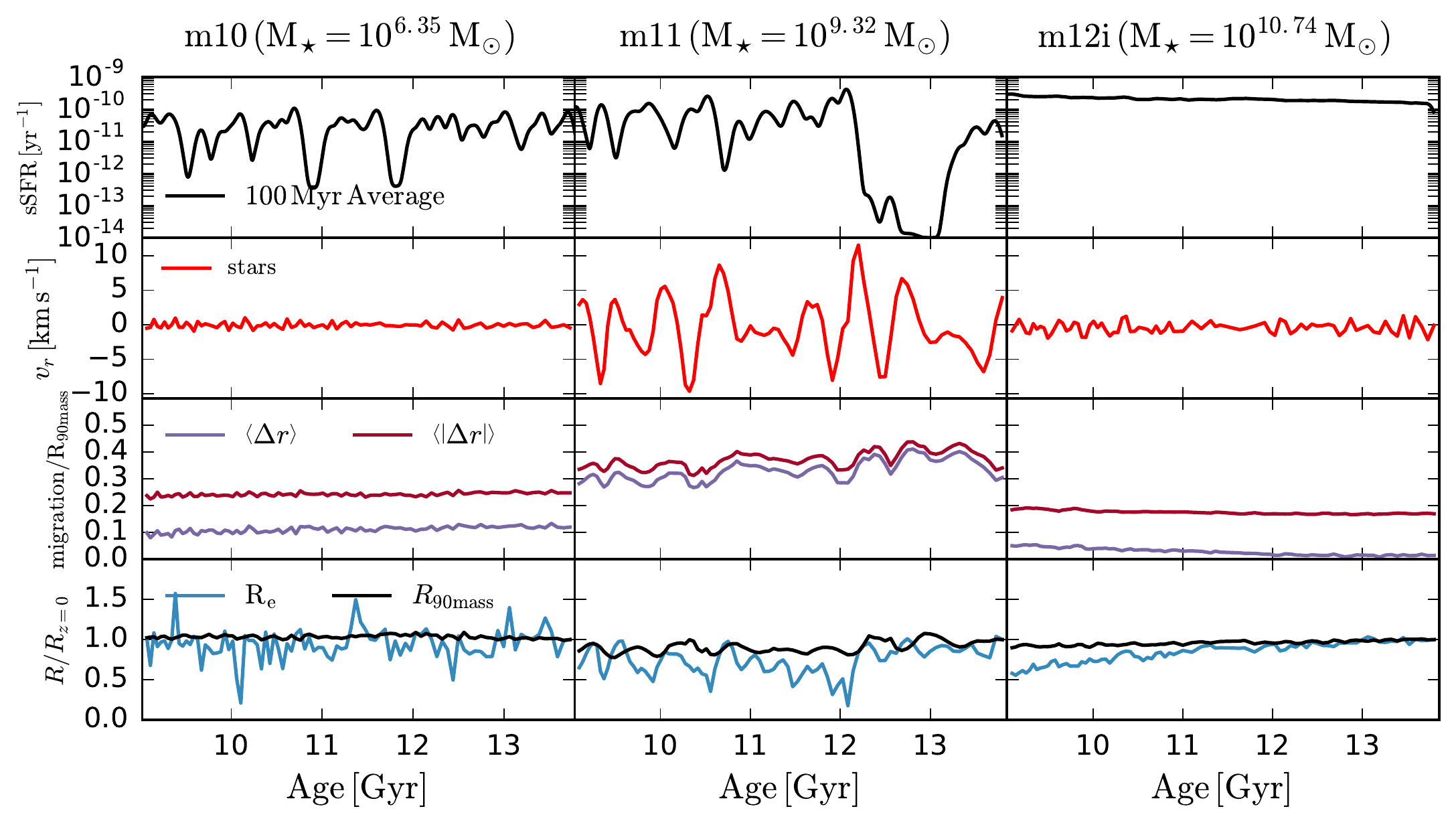}
	\caption{Late-time evolution since $z \approx 0.45$ of galaxies in three mass regimes, as labeled.
\textbf{Row 1}: Specific star formation rate (sSFR) averaged over 100 Myr.
\textbf{Row 2}: Average radial velocity of all stars.
\textbf{Row 3}: Average of the net (purple) and absolute (red) radial migration distance of stars since their formation, scaled to $R_{\rm 90mass}$.
\textbf{Row 4}: Effective radius, $R_e$, which encloses 50\% of the stellar light, and $R_{\rm 90mass}$, which encloses 90\% of $\mstar$, both scaled to their values at $z = 0$.
Star formation is burstier in lower-mass galaxies. However, the effect on stellar kinematics is strongest at $\mstar \sim 10 ^ 9 \msun$. Significantly lower-mass galaxies, such as m10, have more stable kinematics because their overall star-formation efficiency, and thus total feedback, is much lower. This, combined with their lower total baryon fractions, leads to weaker potential fluctuations. However, $R_e$ still fluctuates significantly due to changes in the distribution of the youngest stars, which contribute most of the light. More massive galaxies, such as m12i, also show weaker kinematic fluctuations; their deeper potential wells and lower gas fractions stabilize the potential and inhibit any significant systematic (outward) migration. However, non-trivial absolute (combination of both inward and outward) radial migration still occurs.}
	\label{fig:three_galaxies_long}
\end{figure*}

Figure~\ref{fig:three_galaxies_long} shows the late-time evolution (over the last $\sim5 \gyr$ since $z \sim 0.45$) of three galaxies that span the range of masses in our sample. This shows, as a function of time, the same quantities that we explored above: sSFR (top row), average radial velocity of stars (second row), average radial migration distance of stars (third row), and stellar half-light radius, $R_e$, as well as 90\%-$\mstar$ radius, $R_{\rm 90mass}$ (bottom row).

The sSFRs of the low-mass m10 and m11 are highly bursty and stochastic, with fluctuations similar to m10.6. By contrast, m12i shows much smoother sSFR because (1) it has a much deeper and more stable potential, and (2) being a more massive galaxy, its sSFR is averaged over many more star-forming regions. See \citet{2015arXiv151003869S} for the dependence of star-formation burstiness on mass in our simulations. 

The evolution of stellar radial velocity, radial migration, and size evolution in m11 are all similar to m10.6, with radial velocity fluctuations of $\pm 10 \kms$, typical radial migration since formation of $\approx 5 \kpc$ that fluctuates by $\approx 2 \kpc$ over a few 100 Myr, and $R_e$ that fluctuates by a factor of 2 over a similar timescale.

In m12i, stellar feedback does drive gas out of the disk, but the galaxy's potential well is deeper, with a much smaller contribution from gas in the central regions. Thus, coherent fluctuations in the average radial velocity of stars are limited to a few ${\rm km\,s^{-1}}$. Similarly, the amount of radial migration is much less, though it is non-zero at $1 - 2 \kpc$ on average. As is clear from the lack of late-time fluctuations, most of this migration was seeded at higher redshifts or is the result of scattering processes that affect a smaller fraction of the stars.

Finally, while the sSFR in m10 also is highly bursty, the galaxy's stellar kinematics are relatively stable, with less radial migration. While $R_e$ evolves significantly because of changes in the distribution of young stars, which contribute most of the light, $R_{\rm 90mass}$ remains nearly constant, showing no late-time fluctuations in the distribution of stars. This is because, despite also having shallow potential wells and high gas fractions, galaxies at this mass form too few stars, in part from significant baryonic mass loss early in their evolution from cosmic reionization and stellar feedback \citep{2013ApJ...766...56M, Muratov_2015, 2015arXiv150800007C}. Thus, such galaxies experience insufficient stellar feedback to drive significant gas mass into the halo and significantly change the galactic potential. Bursty star formation alone does not necessarily imply strong stellar kinematic fluctuations and radial migration in galaxies of all masses.

Overall, stellar migration and size evolution are most extreme in galaxies with $\mstar \sim 10 ^ 9 \msun$ (halo $\mthm \sim 10 ^ {11} \msun$). This is the same mass scale where feedback most efficiently produces dark matter cores via the same mechanism \citep[for example,][]{Pontzen_2012, 2015arXiv150703590T, Chan_2015, 2015arXiv150804143R}.

\subsection{Dependence of Radial Migration on Mass}
\label{sec:migration_mass}

We next quantify the amount of stellar radial migration across our mass range, using the procedure described in Section~\ref{sec:measuring_migration}. Again, we compute both $\langle \Delta r \rangle$, the average \textit{net} radial migration, and $\langle \left| \Delta r \right| \rangle$, the average \textit{absolute} radial migration.

\begin{figure}
	\includegraphics[width=\columnwidth]{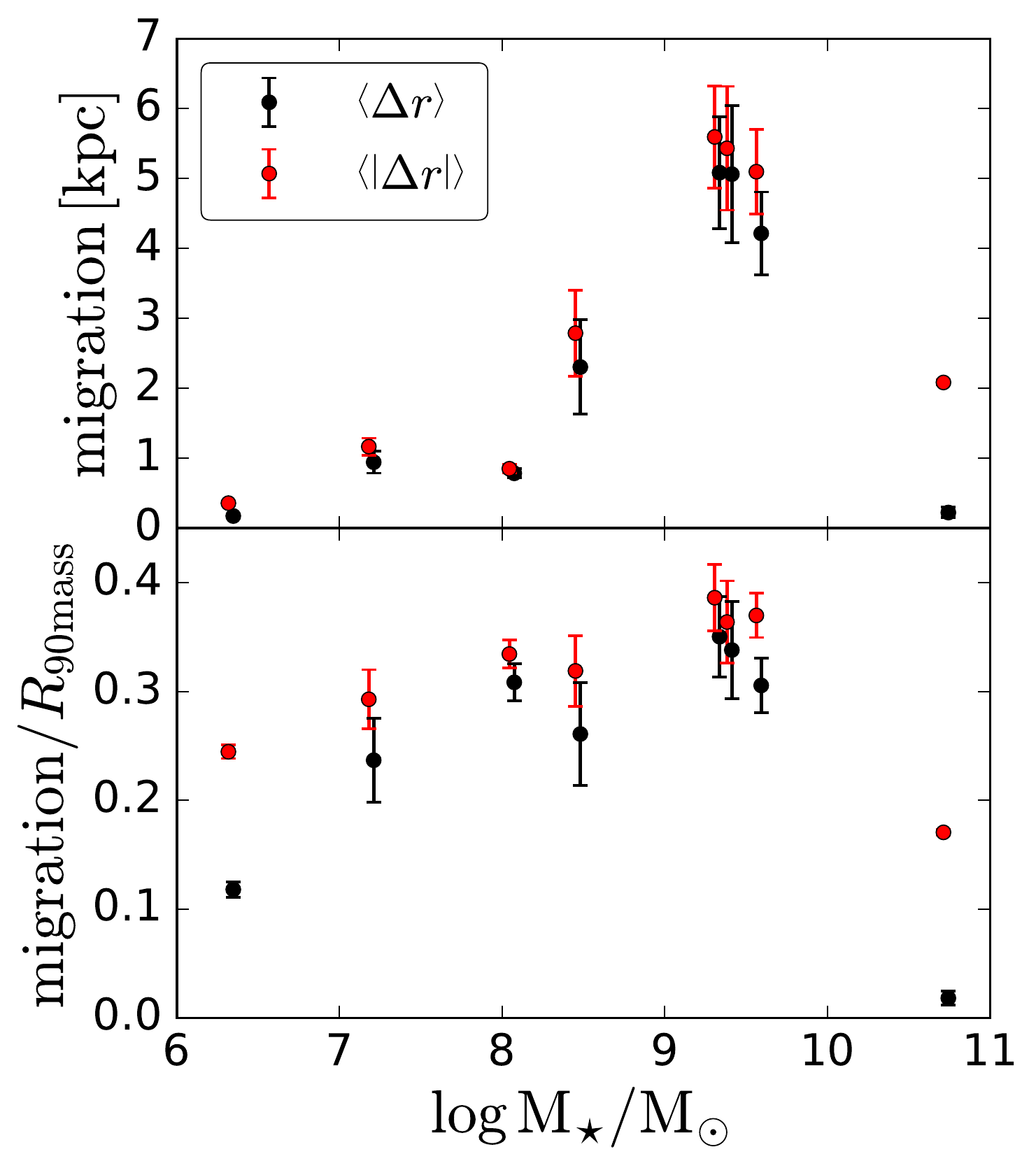}
	\caption{Average radial migration of stars since their formation for all galaxies in our sample. Top panel shows migration in physical kpc, while bottom shows migration scaled by each galaxy's 90\%-$\mstar$ radius, $R_{\rm 90mass}$, at $z = 0$. Black and red points show average of the net and absolute radial distance, respectively, with a small horizontal offset for clarity. For each galaxy, the point shows the average across the 40 snapshots from $z = 0.2$ to $0$, and error bars show the standard deviation across these snapshots, highlighting short-timescale variability. Our highest- and lowest-mass galaxies have little short-time variability (error bars are smaller than points), because their stellar distribution does not change on short timescales. The most significant net outward radial migration occurs in galaxies with $\mstar \approx  10^{7 - 9.6} \msun$. Our highest- and lowest-mass galaxies shows weaker systematic outward migration but do show non-trival absolute radial migration (combination of inward and outward migration), especially as scaled to $R_{\rm 90mass}$. This significant absolute migration of our lowest-mass galaxy is likely the combined result of scattering and stars on radial but stable orbits, since the stellar distribution of these galaxies does not change on short timescales at late times.}
	\label{fig:mig_vs_mass}
\end{figure}

Figure~\ref{fig:mig_vs_mass} shows the average radial migration as a function of $\mstar(z = 0)$. Points show the average migration for each galaxy averaged over the 40 snapshots between $z\sim0.2$ and $0$, and error bars show the standard deviation across these snapshots, highlighting the scatter from short-time variability. The top panel shows migration in physical units, while the bottom panel show the relative amount of migration as scaled to each galaxy's 90\%-$\mstar$ radius, $R_{\rm 90mass}$.

First, the net radial migration, $\langle \Delta r \rangle$ (black points), in either physical units or scaled to $R_{\rm 90mass}$, is largest in galaxies with $\mstar \approx 10 ^ {7 - 9.6} \msun$. This confirms the mass scaling apparent in Figure~\ref{fig:three_galaxies_long}: galaxies with significantly higher or lower mass than this have more stable kinematics (little variability) and little systematic outward migration. Note, however, that a larger sample of simulated galaxies is needed to determine precisely the mass at which migration becomes unimportant. 

Importantly, the galaxies at $\mstar \sim 10 ^ 9 \msun$ with the strongest net radial migration also show the strongest short-time variability (scatter). This is because such short-time variability is required to drive coherent long-term migration: the net migration at late times reflects the permanent dynamical heating of stellar orbits caused by many outflow/inflow episodes that slowly transfer energy to collisionless particles over many Gyr. This is the same phenomenon that is expected to drive dark-matter coring, as we discuss in Section ~\ref{sec:dark_matter_coring}. 

Although m12i has undergone little net migration (black points), it does show non-trivial absolute radial migration (red points) of $\approx 2 \kpc$ on average at $z = 0$. That is, many stars in m12i have migrated away from their formation radius, but approximately equal numbers of stars have migrated inward and outward. Qualitatively, this agrees with previous studies of radial migration in massive disk galaxies, which found that significant radial migration occurs as stars scatter off of massive non-axisymmetric structures such as spiral arms and bars. For example, in a simulation of a disk galaxy with mass similar to m12i, \citet{Ro_kar_2008} found that over the course of 10 Gyr, stars migrated an rms distance of $2.4 \kpc$, with no preference for inward or outward migration.

\subsection{Dependence of Population Gradients on Mass}
\label{sec:mass_dependence_gradients} 

We next examine how the above mass dependence of radial migration affects radial gradients in stellar populations. For each galaxy, we measure the average property (age, total metallicity) of the stellar population at $R_{\rm 90mass}$ and at the galactic center, and we compute the difference across the galaxy such that, for property $P$, $\Delta P = P(r = R_{\rm 90mass}) - P(r = 0)$. We compute this both using stellar radii at $z = 0$ and using each star particle's radius when it formed (also recomputing $R_{\rm 90mass}$ based on formation radii).

\begin{figure}
\includegraphics[width=\columnwidth]{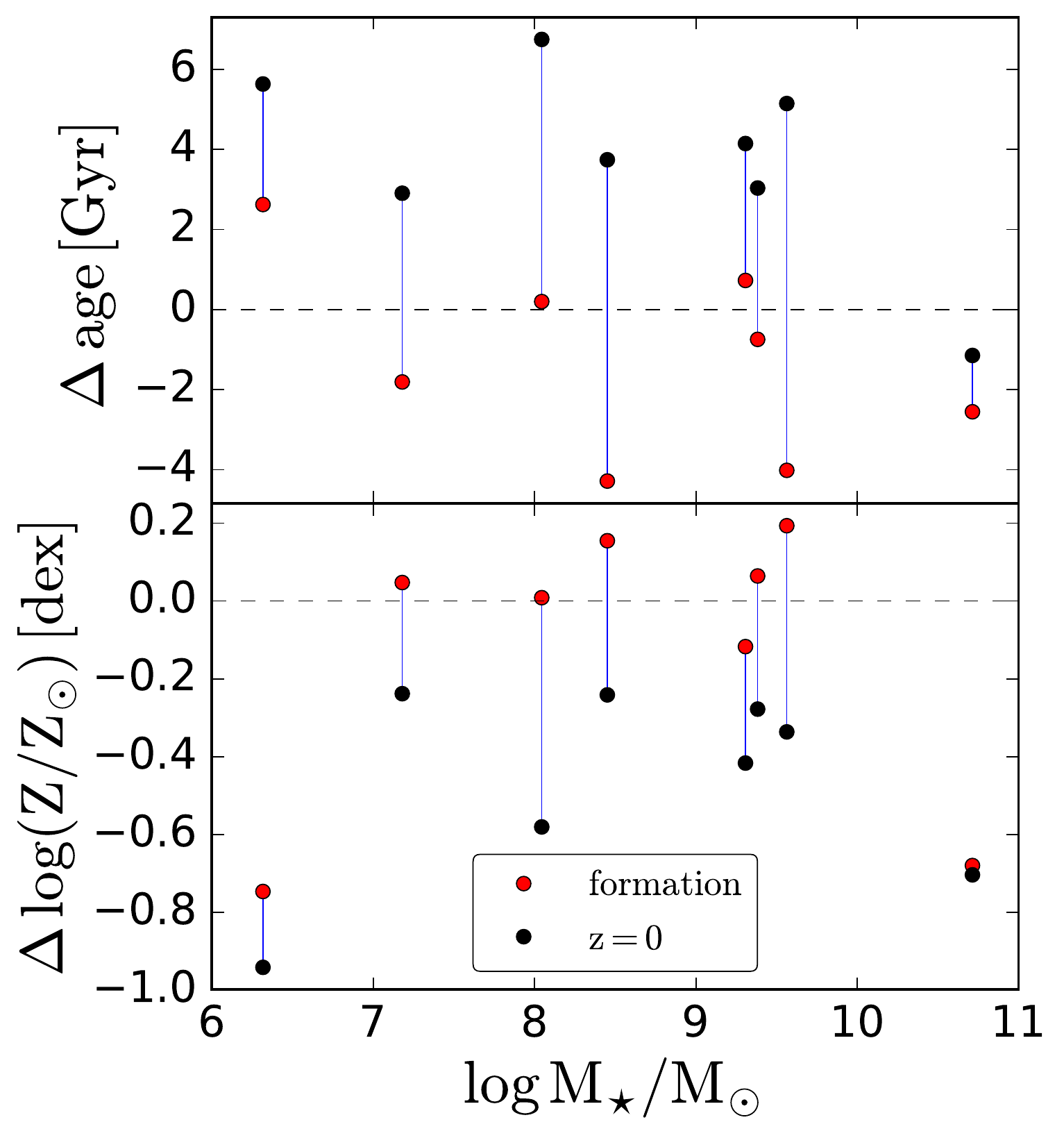}
\caption{Radial gradients in stellar populations for all galaxies in our sample, for average stellar age (top) and total metallicity (bottom).
Black points show the difference between the average age/metallicity of stars at $R_{\rm 90mass}$ and at the galactic center, such that $\Delta P = P(r = R_{\rm 90mass}) - P(r = 0)$, with radii measured at $z = 0$.
Red points show the same but using stars' radius at \textit{formation}, that is, if stars did not migrate from where they formed.
For our lowest- and highest-mass galaxies, where radial migration is weakest, the age gradients remain positive and negative, respectively, such that the gradients at $z = 0$ largely reflect the true underlying gradients at formation.
However, at intermediate masses, the significant outward migration typically inverts the intrinsically negative age gradients at formation, such that they appear positive at $z = 0$.
Metallicity gradients show similar trends: while all galaxies have negative metallicity gradients (more metal-poor stars at larger radii) at $z = 0$, this reflects an inversion of the underlying gradient at formation for most galaxies with $\mstar \approx 10 ^ {7 - 9.6} \msun$.}
		\label{fig:gradients_vs_m_star}
\end{figure}

Figure~\ref{fig:gradients_vs_m_star} shows the gradients in stellar age (top) and total metallicity (bottom). Considering first age, we find strong positive age gradients at $z = 0$ in lower-mass galaxies, with younger stars nearer the galactic center, in qualitative agreement with observations \citep[][and references therein]{Schroyen_2013}. However, at all masses, outward migration since formation has driven these gradients to be more positive (less negative). This effect is weakest in m10 and m12i, which undergo the least net outward migration. m10 is the only galaxy in our sample whose positive age gradient at $z=0$ is not primarily driven by migration. In this galaxy, star formation becomes increasingly centrally concentrated over time; the few star particles which do form at large radius form before $z=2$.

The most dramatic effect occurs in galaxies with $\mstar \approx 10 ^ {7 - 9.6} \msun$. Most of these formed with intrinsically negative age gradients (inside-out growth), but subsequent radial migration has inverted this underlying trend to appear to be positive (appearance of outside-in growth) at $z = 0$. 

Figure~\ref{fig:gradients_vs_m_star} shows similar trends for metallicity gradients. All of our galaxies have negative metallicity gradients at $z = 0$, with more metal-rich stars near the core. For m12i and m10, these gradients are close to the underlying gradients at formation, with modest changes due to subsequent radial migration. However, most of our galaxies with $\mstar \approx 10 ^ {7 - 9.6} \msun$ formed with intrinsically positive (though nearly flat) gradients, such that more metal-rich stars formed at slightly larger radii; migration inverted the true population gradient. 

There is significant variation in the extent to which radial migration alters galaxies’ $z=0$ population gradients, since this depends both on the total amount of migration and on the radial star formation history. Galaxies with extended radial star formation at early times (for example, m11) have positive intrinsic age gradients, which are altered less by migration.

In summary, radial migration can significantly and systematically bias and even invert intrinsic stellar population gradients within galaxies in the critical range of $\mstar \approx 10 ^ {7 - 9.6} \msun$. This means that intrinsic population gradients observed at $z = 0$ do not necessarily reflect the true gradients at formation, thus raising significant concerns for galactic-archaeology studies that aim to infer radial SFHs from stellar populations at $z = 0$.

\subsection{Rapid Size Evolution}
\label{sec:radius_changes}

Finally, we examine size evolution. We have shown that galaxies with $\mstar \approx 10 ^ {7 - 9.6} \msun$ undergo rapid fluctuations that drive significant stellar migration on timescales of a few 100 Myr, which in turn leads to rapid fluctuations of the stellar effective radii by factors of $2 - 3$. We now compare this rapid size evolution to the observed sizes of galaxies at fixed $\mstar$.

This comparison serves two purposes. First, comparing against observations tests whether our simulations produce realistic galaxy sizes, and in particular, whether the dramatic size evolution overestimates the observed scatter in galaxy sizes. Second, the comparison can shed light onto the physical origin of the observed scatter in galaxy sizes at fixed mass.

\begin{figure}
	\includegraphics[width=\columnwidth]{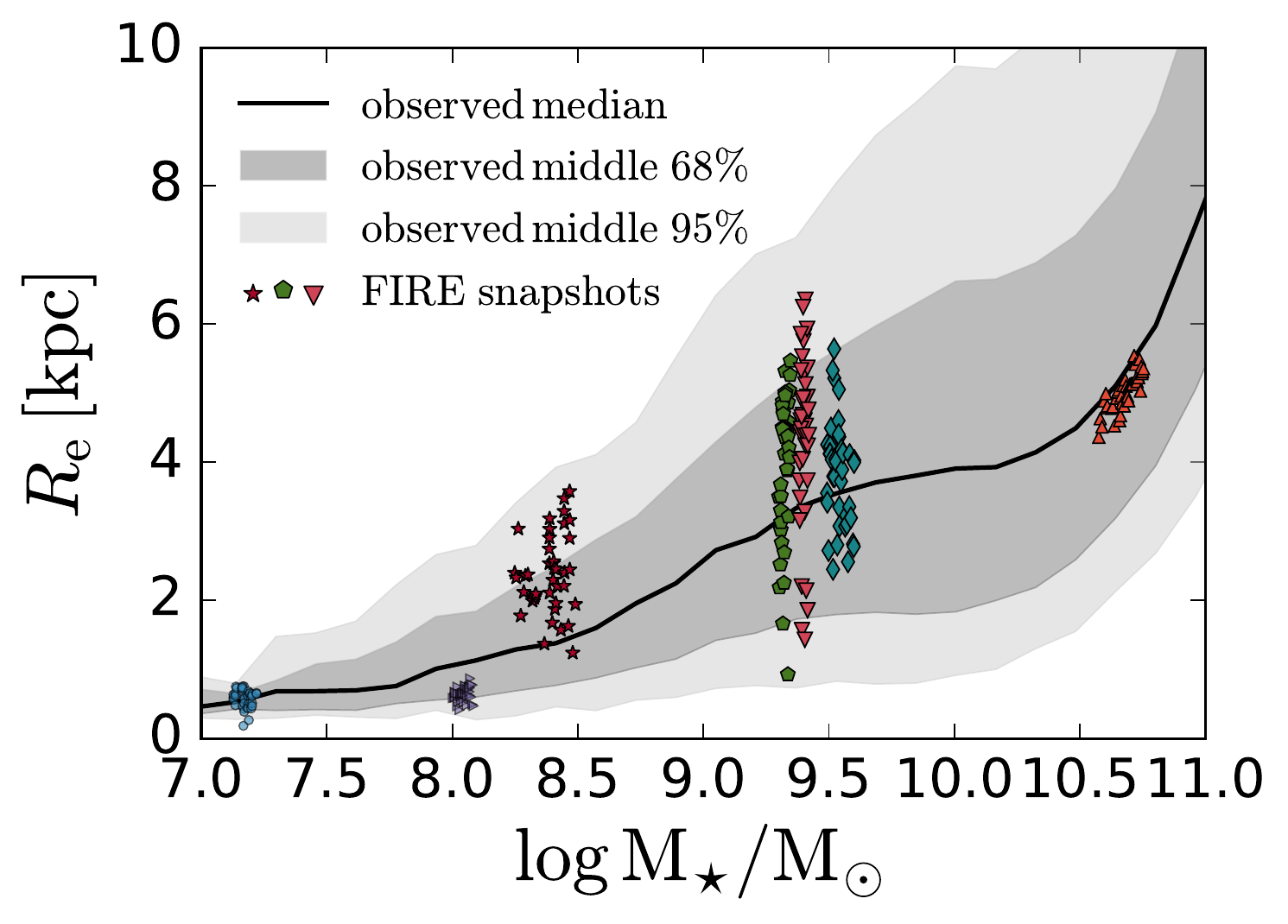}
	\caption{Half-light radius, $R_e$, versus stellar mass, $\mstar$.
Black curve and shaded regions show the median, $1 \sigma$, and $2 \sigma$ scatter for observed isolated galaxies, which we obtain from the NASA-Sloan Atlas (NSA).
Different colored points show 7 of our simulated galaxies (m10.1, m10.2, m10.6, m11, m11v, m11.2 and m12i), sampled across the 40 snapshots from $z = 0.2$ to $0$ (across $\approx 2.5 \gyr$).
The time-averaged sizes of all of our galaxies agree well with observations.
In the critical mass range $\mstar \approx 10 ^ {7 - 9.6} \msun$, feedback-driven outflows cause the radius of an individual galaxy to fluctuate by more than a factor of 2 over just a few 100 Myr. Thus, the short-timescale evolution of individual galaxies can account for much of the observed scatter in radius at fixed $\mstar$.}
	\label{fig:effective_radii} 
\end{figure}

Figure~\ref{fig:effective_radii} shows the SDSS $r$-band half-light radius, $R_e$, as a function of $\mstar$ for galaxies at $z \sim 0$.  We compare the radii measured in our simulations to the effective radii of galaxies observed in the SDSS, as measured in the NASA-Sloan Atlas \citep[NSA;][]{2011AJ....142...31B}. We compare only to galaxies in isolated environments, using the isolated sample described in \citet{Bradford_2015}. For reference, in the lowest $\mstar$ bin in Figure~\ref{fig:effective_radii}, the NSA contains 20 isolated galaxies; all the bins above $10^{8}\msun$ contain at least 100 isolated galaxies. We do not compare against m10 because of the small number of galaxies at $\mstar < 10 ^ {7} \msun$ in the NSA.

The different colored/shaped points show each simulated galaxy sampled across the 40 snapshots from $z = 0.2$ to $0$, highlighting the significant fluctuations in $R_e$ by a factor of $2 - 3$ within just a few 100 Myr. In comparing with observations, we emphasize the importance of computing the half-\textit{light} radius, whose fluctuations are stronger than the half-$\mstar$ radius (not shown). This is because younger stellar populations are brighter and thus disproportionately affect $R_e$, and as we showed in Section~\ref{sec:kinematics}, the kinematic fluctuations and radial migration of young stars are stronger than those of older (fainter) stellar populations.

Figure~\ref{fig:effective_radii} highlights several key results in comparing our simulations with observations. First, the time-averaged $R_e$ of all of these galaxies are consistent with the average $R_e$ of observed galaxies at the same $\mstar$. Thus, even for these highly bursty galaxies, our simulations produce correct sizes. Second, the significant short-timescale fluctuations in $R_e$ for our simulated galaxies remains within the observed scatter. Thus, as dramatic as their size variations are, our simulations do not overpredict the amount of scatter in $R_e$. Finally, and most interestingly, the short-timescale (within just a few 100 Myr) variation in $R_e$ that galaxies with $\mstar \approx 10 ^ {7 - 9.6} \msun$ experience is sufficient to account for a large fraction of the observed scatter in $R_e$ at fixed $\mstar$. This implies that, at least for isolated galaxies at this critical range of $\mstar \approx 10 ^ {7 - 9.6} \msun$, \textit{the observed scatter in radius at fixed $\mstar$ does not simply reflect systematic differences in long-term evolutionary histories, but it also reflects the size fluctuations that individual galaxies undergo within just a few 100 Myr}. 

In the critical mass range, the scatter in the radii of observed galaxies is consistent with being driven primarily by short-timescale fluctuations. These are smaller in m12i, suggesting that the observed scatter at higher masses may be driven primarily by a diversity of long-term evolutionary histories. Given our limited sample, we are unable to determine robustly whether more varied long-term evolutionary histories contribute significant scatter at lower masses.

\section{Comparisons with Previous Work}
\label{sec:comparison}

\subsection{Relation to Dark Matter Core-Creation}
\label{sec:dark_matter_coring}

The transfer of energy from gas outflows/inflows to collisionless particles is well-studied in the context of the ``core-cusp'' \citep{Moore_1994, Oh_2015} and related ``too-big-to-fail'' \citep{Boylan_Kolchin_2011, Jiang_2015} problems, which express the discrepancy between the steep central density profiles (``cusps'') predicted by $\Lambda$CDM models and and the flatter density profiles (``cores'') observed in many nearby low-mass galaxies.
Recently, a number of studies \citep[for example,][]{Read_2005, Pontzen_2012, Governato_2012, Di_Cintio_2013, O_orbe_2015, Chan_2015} have shown that including stellar feedback in simulations can significantly alter dwarf galaxies' density profiles, removing dark matter from the central regions and potentially reconciling the predictions of $\Lambda$CDM models with observations, at least for galaxies with $\mstar = 10 ^ {7 - 9.6} \msun$ ($M_{\rm halo} \approx 10^{10-11.5} \msun$).

We find similar dynamics for the stellar population, which is not surprising given that both stars and dark matter behave as (effectively) collisionless fluids, so that a time-varying gravitational potential will transfer energy to both species, regardless of particle mass \citep{Lynden_Bell_1967, Henriksen_1997, Levin_2008}. Across a wide range of galaxy $\mstar$, the scale lengths of observed and simulated stellar density profiles are approximately equal to the characteristic sizes of galaxies' dark matter cores \citep{Gentile_2009, Governato_2010, Brooks_2011}. This relation also holds true in our simulations: the effective radii $R_e$ of all of our galaxies with $M_{{\rm star}}<10^{10}\msun$ are equal to the galaxies' core radii $r_{\rm core}$ to within a factor of $<2$ \citep[see][]{Chan_2015}. Thus, \textit{if stellar feedback creates dark-matter cores in low-mass galaxies, it also should significantly change the kinematics and spatial distribution of stars}. This connection can be seen clearly in Figure~\ref{fig:sfr_vs_migration}: the time-evolution of the mean stellar migration is very similar to that of $\alpha,$ the central slope of the dark matter density profile.

The mass-scaling of radial migration shown in Figure~\ref{fig:mig_vs_mass} supports this picture, though larger simulated samples are needed to delineate the mass dependence in detail. Both the time-averaged migration and the corresponding scatter are most significant at $\mstar \approx 10 ^ {7-9.6} \msun$ (halo $\mthm \sim 10 ^ {10-11.5} \msun$), with the effect weaker at both higher and lower masses. This scaling is similar to that of dark matter core sizes found by \citet{Chan_2015}, who studied the same simulated galaxies, and both \citet{Di_Cintio_2013} and \citet{2015arXiv150703590T}, who studied larger samples of simulated galaxies in the same mass range.

Galaxies with $\mstar \approx 10^{7-9.6} \msun$ have the optimal balance between shallow gravitational potentials, high gas fractions, sufficiently high star-formation efficiency, and bursty SFRs to allow for significant transfer of energy between gas and collisionless particles. Although higher-mass galaxies have more efficient star formation, their deep gravitational potentials retain most of their gas during periods of high star formation \citep{Muratov_2015, 2015arXiv150800007C, 2015arXiv150703590T}, so they have fewer coherent outflows and more stable SFRs at late cosmic times. Conversely, while galaxies with $\mstar \lesssim 10 ^ {7} \msun$ have bursty SFRs, they have low star-formation efficiency as a result of gas expulsion via cosmic reionization and stellar feedback, so they do not form enough stars to generate the feedback energy needed to significantly change their gravitational potentials at late times. Consistent with this explanation, \citet{Chan_2015} showed that the total energy injected by supernovae alone is sufficient to create dark-matter cores in galaxies with $\mstar \gtrsim 10^9 \msun$, while at significantly lower masses, only small cores can be created even if 100\% of feedback energy is transferred to dark matter.

Radial migration of both stars and dark matter are driven by different dynamics on short and long timescales. On short timescales, migration is semi-periodic and nearly reversible: though stars migrate outward during outflow periods, they migrate back toward the galactic center as gas cools and the potential contracts (see Figures~\ref{fig:sfr_vs_vr} and \ref{fig:sfr_vs_migration}). \citet{Pontzen_2012} showed that such expansion is exactly reversible in the adiabatic limit, when changes in the potential occur on timescales which are long compared to the dynamical time. On the other hand, when outflows change the potential rapidly relative to the dynamical time, energy can be added \textit{permanently} to the orbits of collisionless particles, be they dark matter or stars. Several studies \citep[for example,][]{Pontzen_2012, Ogiya_2014, O_orbe_2015, Chan_2015, 2015arXiv150703590T} found that, while individual outflow episodes temporarily can move dark matter outward on short timescales, many repeated, semi-periodic oscillations are required to excavate lasting cores. Radial migration is strongest for the oldest stars, which have undergone the largest number of inflow-outflow cycles; this generates positive population gradients rather than simply mixing stars of all ages. 

We do, however, note one key difference between the kinematics of stars and dark matter: the migration of young stars can be even stronger on short timescales, because outflowing/inflowing gas can remain star-forming, producing young stars that directly inherit the strong kinematic fluctuations of the feedback-driven gas.

Overall, the similarities between the physical drivers of core creation and stellar migration imply that stellar kinematics and morphologies should provide strong observational tests for baryonic solutions to the ``core-cusp'' and ``too-big-to-fail'' problems, as we will discuss in Section~\ref{sec:observations}.

\subsection{Comparison with Theoretical Work}

Bursty SFHs in low-mass galaxies are characteristic of our FIRE simulations and other simulations that use explicit treatments of stellar feedback at high resolution with a high density threshold for star formation, $n_{\rm SF} \gtrsim 10 \, \mathrm{cm ^ {-3}}$ \citep[for example,][]{2014ApJ...792...99S, 2014ApJ...789L..17M, 2014MNRAS.438.1208K, 2015arXiv151005644H, 2015arXiv151003869S, 2015arXiv151005650H}. \citet{Schroyen_2013} investigated the relation between $n_{\rm SF}$ and the SFHs of low-mass galaxies in idealized simulations including supernova feedback, finding that a high density threshold results in clumpier gas and more clustered star formation as well as increased scattering of stars off gas clumps, leading to moderate stellar migration. Our significant radial migration is consistent with this interpretation, though we find that migration in our simulations is primarily the result of global fluctuations in the potential rather than isolated scattering events.

Using an idealized simulation of a dwarf galaxy, \citet{Teyssier_2013} found that strong supernova feedback leads to the creation of a dark matter core, and this also creates a thick stellar disk that is kinematically hot ($v_{\rm star} / \sigma_{\rm star} \sim 1$), with morphology and kinematics similar to the observed galaxy WLM. Similarly, \citet{2015arXiv151101095W} recently found that isolated dwarf galaxies in the FIRE simulations form as puffy stellar systems that are largely dispersion-supported. They also showed that $\sim 70\%$ of the observed isolated star-forming dwarf galaxies in the Local Group have dispersion-supported (\textit{not} rotation-supported) stellar populations.

Thus, the stellar kinematics of low-mass galaxies in FIRE simulations largely agree with observations in the Local Group. A larger sample of simulations is needed to determine whether the FIRE simulations can also reproduce small fraction of observed rotation-supported low-mass galaxies. It is possible that the FIRE simulations  somewhat overpredict the burstiness of star formation at late times, as \citet{2015arXiv151003869S} found that the fraction of temporarily quenched isolated galaxies at $\mstar < 10 ^ {9} \msun$ in the FIRE simulations is higher (roughly 30\%) than the nearly 0\% observed in nearby isolated galaxies \citep{Geha_2012, Karachentsev_2013}.

Radial migration in low-mass galaxies also has been studied in the context of the formation of stellar halos. \citet{Maxwell_2012} studied a simulated galaxies with $\mstar(z = 0) \sim 5 \times 10 ^ 7 \msun$ and found that, between $z = 8$ and $z = 5$, highly localized and episodic stellar feedback drove rapid gas flows that in turn drove stellar kinematics in the central $\sim 100 \pc$. This caused stars that formed in the central regions to migrate outward significantly. However, they found that migration became significantly less efficient by $z = 5$ and predicted that it would become negligible at late times. Similarly, \citet{Stinson_2009} found significant radial migration in the oldest stellar populations of simulated galaxies with $\mstar(z = 0) = 10 ^ {6 - 8} \msun$, suggesting that radial migration, rather than accretion events, might be responsible for the creation of the stellar halo. Consistent with this result, Figure~\ref{fig:all_migration} showed that the oldest stars ($> 8 \gyr$) in m10.6 have migrated outward an average of $\sim 4 \kpc$, or $2\,R_e(z= 0)$, from the radius where they formed; many of these stars likely would be classified as part of the stellar halo.

Finally, \citet{2015arXiv151106188B} recently studied the origin of observed population gradients in simulated dwarf galaxies using the CLUES cosmological zoom-in simulation of a Local Group analogue. They found that the observed positive gradients of stellar age with radius in dwarf galaxies can be generated via major mergers, which heat old stellar populations and trigger new bursts of centrally concentrated star formation. Our results are consistent with this possibility; however, the low-mass galaxies that we study have relatively calm merger histories, and despite this, almost all of them form positive age gradients. Thus, our results imply that positive gradients in stellar age can form even in the absence of late-time mergers, via stellar feedback and bursty star formation.

\subsection{Comparison with Observations}

Low-mass galaxies in the local Universe exhibit significant scatter in their SFRs, consistent with high variability on short timescales, and many show evidence for multiple episodes of star formation separated by periods of quiescence ranging from a few 10's of Myr to several Gyr \citep[for example,][]{2003AJ....126..187D, 2004MmSAI..75..110R, 2005NewAR..49..453S, 2013A&A...549A..47L, 2014A&A...572A..10D, Weisz2014, 2015ApJ...805..103V, Geha_2015}. Most of these star-forming galaxies are relatively isolated at $z = 0$, indicating that starbursts are triggered not just by interactions. \citet{2009ApJ...692.1305L} found that $6^{+4}_{-2}$ percent of $L \lesssim 0.1 L_{\star}$ galaxies in the local volume are currently undergoing a starburst. \citet{Weisz2012} used $\halpha$-to-$UV$ flux ratios to constrain bursty star formation in 185 galaxies, finding that low-mass galaxies are best described by SFHs with burst amplitudes of $\sim 30$ and interburst spacings of $\sim 150 – 200 \myr.$ \citet{2015arXiv151003869S} compared galaxies from the FIRE simulations to observations using the same $\halpha$-to-$UV$ ratios and found that burst spacings and amplitudes from the simulations were broadly consistent with observationally-inferred values, though the FIRE simulations may somewhat overpredict burst amplitudes at lower masses. 

Almost all low-mass galaxies in the local Universe have negative metallicity gradients. Several studies \citep[for example,][]{Mehlert_2003, Spolaor_2009, Kirby_2011, Schroyen_2013, Vargas_2015, Pilyugin_2015} find an average metallicity gradient of $\Delta \mathrm{[Fe/H]\sim -0.4\,dex}$ between the galactic center and the effective radius, but with significant ($\sim 0.3\,\mathrm{dex}$) scatter even for galaxies with similar morphologies and SFHs. These negative metallicity gradients are broadly consistent with our results (see Ma et al., in prep., for more details). Figure~\ref{fig:gradients_vs_m_star} shows that in our simulations, radial migration plays a critical role in establishing these gradients: without migration, the age and metallicity gradients of galaxies with $\mstar\approx10 ^ {7 - 9.6}\msun$ would be nearly 0. 

\section{Summary and Discussion}
\label{sec:summary}

\subsection{Summary}

Using the FIRE suite of cosmological zoom-in hydrodynamic simulations of isolated low-mass galaxies across $\mstar(z = 0) = 2 \times 10 ^ 6 - 5 \times 10 ^ {10} \msun$, we have explored the effects of stellar feedback and bursty star formation on stellar kinematics, radial migration, size evolution, and population gradients at late cosmic times, where such low-mass galaxies are observable. In galaxies with $\mstar \approx 10 ^ {7 - 9.6} \msun$, stellar feedback frequently drives outflows of significant gas mass well beyond the stellar radius; these cool and fall back into the galaxy center on timescales of a few 100 Myr. These outflow/infall cycles occur semi-periodically many times throughout these galaxies' evolutionary histories and drive strong fluctuations in the galactic potential, leading to dramatic effects on the stellar population. We summarize our main results as follows.

\begin{enumerate}[leftmargin=*]
\item \textit{Physical origin of stellar migration}:
Stars in low-mass galaxies experience significant radial migration via two related processes. First, outflowing/inflowing gas clouds can remain star-forming, producing young stars that inherit the outflowing/infalling kinematics of gas. Second, gas outflows/inflows drive strong fluctuations in the overall galactic potential, and in response, the orbits of stars of all ages expand and contract over short timescales and gradually are heated over long timescales. These physical processes are fundamentally different from the scattering processes known to cause radial migration in massive disk galaxies.

\item \textit{Timescales of stellar migration}:
Stellar radial migration occurs over both short (a few 100 Myr) and long (a few Gyr) timescales. Short-timescale fluctuations in the potential result in rapid changes in kinematics and radial migration for all stars within a few 100 Myr, comparable to the galaxy's dynamical time. Young stars experience the strongest short-timescale migration, typically $\gtrsim 1 \kpc$ within their first $200 \myr$, because they are most tightly coupled to gas kinematics, but even old stars migrate significantly on short timescales. Short-timescale migration, however, is nearly reversible. On the other hand, the \textit{cumulative} effects of many repeated semi-periodic fluctuations gradually heat stellar orbits, driving \textit{permanent} and coherent outward migration of stars over Gyr timescales. Thus, the amount of stellar migration depends on stellar age: the oldest stars, having experienced the most outflow/inflow cycles, exhibit the strongest systematic outward migration since formation.

\item \textit{Impact on radial gradients of stellar populations}:
Stellar migration can systematically  change radial gradients of stellar age and metallicity because the amount of outward migration depends on stellar age. For almost all of our galaxies at $\mstar \approx 10 ^ {7 - 9.6} \msun$, stellar migration inverts true underlying age gradients from negative to positive, or metallicity gradients from positive to negative, by $z = 0$. This means that population gradients observed at $z = 0$ do not necessarily reflect the intrinsic gradients at formation, and that radial SFHs inferred from present-day populations gradients, as common in galactic-archaeology studies, may be significantly biased.

\item \textit{Fluctuations of galaxy sizes}:
Our simulations produce consistent half-light radii ($R_e$) of galaxies as compared with observations. Short-term stellar migration leads to fluctuations in effective radius by factors of $2 - 3$ within just a few 100 Myr. These fluctuations are comparable to the observed scatter in $R_e$ at fixed $\mstar$, which suggests that the observed scatter does not just reflect systematic differences in long-term evolutionary histories, but it also reflects the size fluctuations that individual galaxies undergo within just a few 100 Myr.

\item \textit{Dependence on galaxy mass}:
All of these effects are strongest in galaxies with $\mstar(z = 0) \approx 10 ^ {7 - 9.6} \msun$ (halo $M_{\rm halo} \sim 10 ^ {10-11.5} \msun$). This is the same mass range where stellar feedback has been shown to most efficiently produce dark matter cores. Since migration and coring are driven by the same physical processes, galaxies with significant feedback-driven cores also should have experienced significant stellar migration. Although galaxies with lower mass also experience similarly bursty SFHs, their higher dark-matter fractions and lower overall SFRs (in particular, low $\sfr / M_{\rm halo}$) lead to weaker fluctuations in the galactic potential, with weak effects on stellar kinematics and radial migration. More massive galaxies have deeper potential wells and lower gas fractions within $R_e$, so they do not experience strong coherent fluctuations in stellar kinematics.

\item \textit{Stellar kinematics provide a strong test to constrain feedback models}:
Our stellar feedback models predict strong effects on stellar kinematics and sizes of low-mass galaxies.
If stellar feedback drives dark-matter coring, galaxies with large cores also should have experienced significant stellar migration.
Detailed studies of nearby dwarf galaxies, with resolved spectroscopy and/or proper motions of individual stars, therefore can provide strong tests of our model predictions, as we outline below.
\end{enumerate}

\subsection{Discussion: Implications for Observational Tests of Stellar Feedback}
\label{sec:observations}

We have shown that stellar feedback can cause dramatic changes in the radial kinematics and distribution of stars in isolated galaxies with $\mstar \approx 10 ^ {7 - 9.6} \msun$. These effects are analogous to feedback-driven coring of such galaxies' dark-matter profiles, as explored in previous works. Because the distribution and kinematics of stars (and gas) are more directly observable than (inferred) dark-matter mass profiles, our results imply that detailed observations of resolved stellar populations and kinematics, which are obtainable for nearby galaxies, can provide strong tests of stellar feedback models. We outline a few possible observational tests, which we will investigate in more detail in future work.

\textit{Biased stellar kinematics}:
Figure~\ref{fig:sfr_vs_vr} (bottom) showed that because young stars inherit the kinematics of outflowing/infalling gas from which they form, young stars can have biased kinematics as compared with older stars. In addition, Figure~\ref{fig:sfr_vs_vr} suggested that the observed radial velocity (or line-of-sight velocity dispersion) should be higher in actively star-forming galaxies than in post-starburst galaxies with extended outflows. Indeed, we find that the velocity dispersion falls by $\sim 40\%$ during outflow periods, when stars are more weakly bound. Finally, at any stage of evolution, our results suggest that the orbits of stars in low-mass galaxies may be strongly anisotropic, in particular, strongly radially biased.

\textit{Gas outflows}:
Figure~\ref{fig:gas_star_distribution} showed that significant neutral gas is blown beyond the outskirts of the stellar distribution following starburst episodes. First, these outflows should be observable directly in nearby post-starburst galaxies. Several of the late-time outflows in m10.6, m11, m11v, and m11.2 carry more than $10 ^ 8 \msun$ of neutral hydrogen beyond $R_{\rm 90mass}$, which is at least $10 \times$ more than the lower limits of resolved HI interferometric observations of nearby galaxies \citep{2012AJ....144..134H}. Second, one could compare the short-timescale fluctuations in HI mass in our simulated galaxies over a single starburst cycle to the scatter in observed HI mass at fixed $\mstar$ \citep[see, for example,][]{Bradford_2015}.

\textit{Relation between population gradients and cores}:
Figure~\ref{fig:gradients_vs_m_star} showed that stellar feedback dramatically can affect and even invert age and metallicity radial gradients in low-mass galaxies. Although our simulated galaxies experience inside-out growth, radial migration erases or inverts these population gradients because old stars migrate outward more than young stars. Because the physical processes responsible for stellar migration are the same processes that produce dark matter cores, our results appear to imply that, at fixed $\mstar$, galaxies with stronger cores will have more positive age gradients and more negative metallicity gradients than those with cusps or weaker cores. However, a larger suite of simulations, with a greater diversity of late-time population gradients, is needed to make this prediction concrete, since all the low-mass galaxies we study here develop both positive age gradients and strong dark matter cores.

\textit{Stellar halos of low-mass galaxies}:
Figure~\ref{fig:all_migration} showed that the oldest stars ($> 8 \gyr$) in m10.6 have migrated outward the most, an average of $\sim 4 \kpc$, or $2\,R_e(z=0)$, from where they formed, and Figure~\ref{fig:now_formation} showed that the oldest stars (on average) are at the largest radii at $z = 0$. Many of these oldest stars migrated sufficiently far that they likely would be classified as part of the stellar halo. Thus, our results suggest that the stellar halos of such galaxies may provide observational probes of the earliest stars that formed near the core, as has also been suggested by studies that find radial migration at high redshift \citep{Stinson_2009, Maxwell_2012}.

Our analysis represents an early examination into stellar kinematics and radial migration in low-mass galaxies. These results motivate more comprehensive theoretical studies, with larger simulation samples, to delineate further the dependence on mass and formation history, as well as the scatter across populations.

%\section{Acknowledgments}
\vspace{5mm} 

We thank Jeremy Bradford for sharing observational data and Coral Wheeler and Frank van den Bosch for useful discussions and comments. We also thank the reviewer for helpful comments.
KE acknowledges support from the Caltech SURF program.
ARW gratefully acknowledges support from the Moore Center for Theoretical Cosmology and Physics at Caltech via a Moore Prize Fellowship, and from Carnegie Observatories via a Carnegie Fellowship in Theoretical Astrophysics.
MG acknowledges a fellowship from the John S. Guggenheim Memorial Foundation.
Support for PFH was provided by an Alfred P. Sloan Research Fellowship, NASA ATP Grant NNX14AH35G, and NSF Collaborative Research Grant \#1411920 and CAREER grant \#1455342.
DK and TKC were supported in part by NSF grant AST-1412153, and funds from the University of California San Diego.  
CAFG was supported by NSF through grants AST-1412836 and AST-1517491, by NASA through grant NNX15AB22G, and by Northwestern University funds. 
Numerical calculations were run on the Caltech compute cluster ``Zwicky'' (NSF MRI award \#PHY-0960291) and on allocations TG-AST120025 and TG-AST130039 granted by the Extreme Science and Engineering Discovery Environment (XSEDE) supported by the NSF.

\end{document}